\documentclass[fleqn,usenatbib]{mnras}
\usepackage{newtxtext,newtxmath}

\usepackage[T1]{fontenc}
\usepackage{ae,aecompl}
\usepackage{tabularx}
\usepackage{epsfig}
\usepackage{graphicx}
\usepackage{amsmath}
\usepackage{amssymb}
\usepackage{mathrsfs}
\usepackage{mathtools}
\usepackage{hyperref}
\usepackage{afterpage}
\usepackage{epsf} 

\usepackage{graphicx}
\usepackage[colorinlistoftodos]{todonotes}

\defcitealias{LVCGCNS190408an}{24096}
\defcitealias{LVCGCNS190412m}{24098}
\defcitealias{LVC2GCNS190421ar}{24141}
\defcitealias{LVCGCNS190425z}{24168}
\defcitealias{LVCGCNS190426c}{24237}
\defcitealias{LVCGCNS190510g}{24442}
\defcitealias{LVCGCNS190512at}{24503}
\defcitealias{LVCGCNS190513bm}{24522}
\defcitealias{LVCGCNS190517h}{24570}
\defcitealias{LVCGCNS190519bj}{24598}
\defcitealias{LVCGCNS190521g}{24621}
\defcitealias{LVCGCNS190521r}{24632}
\defcitealias{LVCGCNS190630ag}{24922}
\defcitealias{LVCGCNS190706ai}{24998}
\defcitealias{LVCGCNS190707q}{25012}
\defcitealias{LVCGCNS190718y}{25087}
\defcitealias{LVCGCNS190720a}{25115}
\defcitealias{LVCGCNS190727h}{25164}
\defcitealias{LVCGCNS190728q}{25187}
\defcitealias{LVCGCNS190814bv}{25324}
\defcitealias{LVCGCNS190828j}{25497}
\defcitealias{LVCGCNS190828l}{25503}
\defcitealias{LVCGCNS190901ap}{25606}
\defcitealias{LVCGCNS190910d}{25695}
\defcitealias{LVCGCNS190915ak}{25753}
\defcitealias{LVCGCNS190923y}{25814}
\defcitealias{LVCGCNS190924h}{25829}
\defcitealias{LVCGCNS190930s}{25871}
\defcitealias{LVCGCNS190930t}{25876}

\newcommand{\cosmo}{$H_0 = 67.4$~km~s$^{-1}$~Mpc$^{-1}$ $\Omega_M = 0.315$ and $\Omega_{\Lambda} = 0.685$ \citep{PlanckCollaboration18}}

\title[Searching for EM Counterparts to GW Merger Events with GOTO-4]{Searching for Electromagnetic Counterparts to Gravitational-wave Merger Events with the Prototype Gravitational-wave Optical Transient Observer (GOTO-4)}

\author[B. P. Gompertz\ et al.]{\href{https://orcid.org/0000-0002-5826-0548}{\includegraphics[scale=0.05]{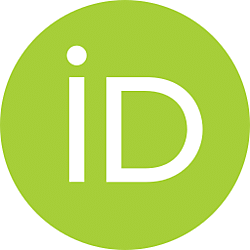}}B. P. Gompertz,\thanks{E-mail: b.gompertz@warwick.ac.uk}$^{1}$
\href{https://orcid.org/0000-0001-8945-5551}{\includegraphics[scale=0.05]{figs/ID.png}}R. Cutter,\thanks{E-mail: r.cutter@warwick.ac.uk}$^{1}$   
\href{https://orcid.org/0000-0003-0771-4746}{\includegraphics[scale=0.05]{figs/ID.png}}D. Steeghs,$^{1,3}$
\href{https://orcid.org/0000-0002-6558-5121}{\includegraphics[scale=0.05]{figs/ID.png}}D. K. Galloway,$^{2,3}$
J. Lyman,$^{1}$
\newauthor
\href{https://orcid.org/0000-0001-6364-408X}{\includegraphics[scale=0.05]{figs/ID.png}}
K. Ulaczyk,$^{1}$
\href{https://orcid.org/0000-0003-3665-5482}{\includegraphics[scale=0.05]{figs/ID.png}}
M. J. Dyer,$^{4}$
K. Ackley,$^{2,3}$
V. S. Dhillon,$^{4,5}$
P. T. O'Brien,$^{6}$
\newauthor
G. Ramsay,$^{7}$
S. Poshyachinda,$^{8}$
\href{https://orcid.org/0000-0001-5455-3653}{\includegraphics[scale=0.05]{figs/ID.png}}
R. Kotak,$^{9}$
L. Nuttall,$^{10}$
R. P. Breton,$^{11}$
E. Pall\'e,$^{5}$
\newauthor
D. Pollacco,$^{1}$
E. Thrane,$^{2}$
S. Aukkaravittayapun,$^{8}$
S. Awiphan,$^{8}$
M. J. I. Brown,$^{2}$
\newauthor
U. Burhanudin,$^{4}$
P. Chote,$^{1}$
A. A. Chrimes,$^{1}$
E. Daw,$^{4}$
C. Duffy,$^{7}$
\newauthor
R. A. J. Eyles-Ferris,$^{6}$
T. Heikkil\"a,$^{9}$
P. Irawati,$^8$
M. R. Kennedy,$^{11}$
T. Killestein,$^{1}$
\newauthor
A. J. Levan,$^{1,12}$
S. Littlefair,$^{4}$
L. Makrygianni,$^{4}$
T. Marsh,$^{1}$
D. Mata S\'{a}nchez,$^{11}$
\newauthor
S. Mattila,$^{9}$
J. Maund,$^{4}$
J. McCormac,$^{1}$
D. Mkrtichian,$^8$
Y. -L. Mong,$^{2,3}$
J. Mullaney,$^{4}$
\newauthor
B. M\"uller,$^{2,3}$
A. Obradovic,$^{2}$
E. Rol,$^{2}$
U. Sawangwit,$^8$
E. R. Stanway,$^{1}$
\newauthor
R. L. C. Starling,$^{6}$
P. A. Str\o m,$^{1}$
S. Tooke,$^{6}$
R. West,$^{1}$
K. Wiersema$^{1,6}$
\\
\vspace{0.1in}\\ 
$^{1}$ Department of Physics, University of Warwick, Gibbet Hill Road, Coventry, CV4 7AL, UK \\
$^{2}$ School of Physics \& Astronomy, Monash University, Clayton VIC 3800, Australia \\
$^{3}$ OzGRav-Monash, Monash University, Clayton VIC 3800, Australia \\
$^{4}$ Department of Physics and Astronomy, University of Sheffield, Sheffield, S3 7RH, UK\\
$^{5}$ Instituto de Astrof\'{i}sica de Canarias (IAC), E-38205 La Laguna, Tenerife, Spain\\
$^{6}$ School of Physics \& Astronomy, University of Leicester, University Road, Leicester LE1 7RH, UK\\
$^{7}$ Armagh Observatory \& Planetarium, College Hill, Armagh, BT61 9DG\\
$^{8}$ National Astronomical Research Institute of Thailand, 260  Moo 4, T. Donkaew,  A. Maerim, Chiangmai, 50180, Thailand\\
$^{9}$ Department of Physics and Astronomy, University of Turku, FI-20014, Turku, Finland\\
$^{10}$ University of Portsmouth, Portsmouth, PO1 3FX, UK\\
$^{11}$ Jodrell Bank Centre for Astrophysics, Department of Physics and Astronomy, The University of Manchester, Manchester M13 9PL, UK\\
$^{12}$ Department of Astrophysics/IMAPP, Radboud University, PO Box 9010, NL-6500 GL Nijmegen, The Netherlands\\
}

\date{Received YYY; in original form ZZZ}

\begin{document}
\defcitealias{US}{H19}
\maketitle
\newpage
\begin{abstract}
We report the results of optical follow-up observations of 29 gravitational-wave triggers during the first half of the LIGO-Virgo Collaboration (LVC) O3 run with the Gravitational-wave Optical Transient Observer (GOTO) in its prototype 4-telescope configuration (GOTO-4). While no viable electromagnetic counterpart candidate was identified, we estimate our 3D (volumetric) coverage using test light curves of on- and off-axis gamma-ray bursts and kilonovae. In cases where the source region was observable immediately, GOTO-4 was able to respond to a GW alert in less than a minute. The average time of first observation was $8.79$ hours after receiving an alert ($9.90$ hours after trigger). A mean of $732.3$ square degrees were tiled per event, representing on average $45.3$ per cent of the LVC probability map, or $70.3$ per cent of the observable probability. This coverage will further improve as the facility scales up alongside the localisation performance of the evolving gravitational-wave detector network. Even in its 4-telescope prototype configuration, GOTO is capable of detecting AT2017gfo-like kilonovae beyond 200~Mpc in favourable observing conditions. We cannot currently place meaningful electromagnetic limits on the population of distant ($\hat{D}_L = 1.3$~Gpc) binary black hole mergers because our test models are too faint to recover at this distance. However, as GOTO is upgraded towards its full 32-telescope, 2 node (La Palma \& Australia) configuration, it is expected to be sufficiently sensitive to cover the predicted O4 binary neutron star merger volume, and will be able to respond to both northern and southern triggers.
\end{abstract}


\section{Introduction}
Gravitational-Wave (GW) detections of compact object mergers are fast becoming common occurrences. During the first half of the advanced LIGO (aLIGO) - advanced Virgo (AdV) Collaboration (LVC) observing run 3 (O3a), which ran from April the 1$^{\rm st}$ to September the 30$^{\rm th}$ 2019, 33 candidate compact object merger events were reported, with 21 of these classified as most likely due to Binary Black Hole (BBH) mergers. The tantalising prospect of finding an electromagnetic (EM) counterpart to GW events warrants follow-up, particularly in light of the rich scientific yield attained following the detection of both GW and EM signals from the Binary Neutron Star (BNS) merger GW170817 \citep{abbott2017multi}. The dominant demographic of GW events, BBH mergers, are expected to be EM silent \citep{abbott2016localization}, although numerous studies suggest that there is potential for an EM counterpart \citep{palenzuela2010understanding,moesta2012detectability,loeb2016electromagnetic,murase2016ultrafast,perna2016short,de2017electromagnetic,janiuk2017possible, Chang18,McKernan19}, and a possible high-energy detection was claimed alongside GW150914 \citep{connaughton2016fermi}. Assessing observational constraints on the BBH merger population has value, because it helps to inform the follow-up strategy of ``wide-fast'' survey telescopes chasing GW triggers, and because constraints on the EM signal will naturally become more stringent as the sample increases. Furthermore, there is not yet sufficient observational evidence to rule out the possibility of an EM counterpart in BBH mergers.

More promisingly, the O3a GW trigger population also contains candidate BNS and Neutron Star-Black Hole (NSBH) mergers, which are expected to exhibit an EM signature, at least in some cases. The prime candidate for such a counterpart is a kilonova \citep[KN;][]{Li98,Rosswog05,metzger2010electromagnetic,Barnes13,Metzger17}, where heavy unstable elements are formed in the neutron-rich environment of the merger via rapid capture (r-process) nucleosynthesis \citep{Lattimer74,Eichler89,Freiburghaus99}, and subsequently produce thermal emission as they decay radioactively. We may also expect to observe a short duration \citep[$< 2$s;][]{Kouveliotou93} gamma-ray burst \citep[sGRB;][]{Blinnikov84,paczynski1986gamma,Eichler89,Narayan92,Rosswog03,belczynski2006study,Fong13}, wherein the merger launches a relativistic jet that can be detected at high energies, and subsequently produces a broadband synchrotron afterglow as the ejecta decelerate and form shocks in the ambient environment \citep{Blandford76}. The KN signature from an NSBH merger is expected to be different to those produced by a BNS \citep{Kawaguchi16,Tanaka18,Barbieri19} -- typically predicted to be brighter in the infrared \citep[e.g.][]{Metzger17,Kawaguchi20}, although see \citet{Foucart19}. There is also some evidence to suggest that NSBH mergers may be distinguishable in the observed sGRB population \citep[e.g.][]{Troja08,Gompertz20}.

 KNe and sGRBs are already known to be linked to one another through coincident detections \citep{Berger13,Tanvir13,Yang15,Jin16,Kasliwal17,gompertz2018diversity,Jin18,Troja18,Jin20,Lamb19b,Troja19,Jin20}, with both BNS and NSBH KN models employed. KNe and sGRBs were also confirmed to be linked to BNS mergers by the detections of GRB 170817A \citep{abbott2017multi,Goldstein17,Hallinan17,Margutti17,Savchenko17,Troja17b,Margutti18,Mooley18,Troja18b,DAvanzo18,Lyman18} and the associated kilonova AT2017gfo \citep{Blanchard17,Chornock17,Coulter17a,Covino17,Cowperthwaite17,Drout17,Evans17,Hjorth17,Levan17,Nicholl17,Pian17,Smartt17,Soares-Santos17,Tanvir17,Villar17} alongside GW170817 \citep{Abbott17b}. In this instance, the sGRB afterglow was likely viewed somewhat away from the jet axis \citep{Abbott17b,Haggard17,Kim17,Lazzati17,Lamb18,Mandel18,Fong19,Lamb19,Wu19}.

Locating EM counterparts to GW detections remains an extremely challenging task due to the very large localisation uncertainties from the GW detectors. These  uncertainties can span from hundreds to tens of thousands of square degrees, and while the localisation performance will improve as the detector network expands \citep{Abbott20}, wide-field survey telescopes remain essential when searching for associated transients. The Gravitational-wave Optical Transient Observer (GOTO; see Steeghs et al., in prep; \citealt{dyer2018telescope}) is one such facility. At design specifications, the project will include one northern node on La Palma (Spain), and one southern node (Australia), each of which will be equipped with 16 telescopes on two robotic mounts, with a field of view of $\sim75$ square degrees per node. The facility will provide rapid-response tiling of the large LVC error boxes backed by an ongoing sky patrol survey. By exploiting its high cadence and wide field, GOTO can quickly identify potential GW counterparts and flag them for further photometric and spectroscopic follow-up. During O3a, GOTO consisted of four wide-field-of-view telescopes, each covering 4.8 square degrees, with a combined footprint of $\approx19$ square degrees. GOTO's first installation is located at the Observatorio del Roque de los Muchachos on La Palma and is able to scan the visible Northern hemisphere sky once every $\sim$ 14 days. Hereafter, ``GOTO-4'' specifically refers to the 4-telescope prototype configuration.

Outside of GOTO, there has been a widespread, sustained effort by the observational community to identify EM counterparts to GW triggers throughout the LVC O3 observing run. Numerous wide-field follow-up missions have tiled GW error boxes searching for transients, including the All-Sky Automated Survery for Supernovae \citep[ASASSN;][]{Shappee14}, the Asteroid Terrestrial impact Last Alert System \citep[ATLAS;][]{Tonry18}, the Deca-Degree Optical Transient Imager \citep[DDOTI;][]{Watson16}, the Dark Energy Survey \citep[DES;][]{DES16}, the Global Rapid Advanced Network Devoted to the Multi-messenger Addicts \citep[GRANDMA;][]{Antier20}, KMT-Net \citep{Kim16}, the Mobile Astronomical System of TElescope Robots \citep[MASTER;][]{Lipunov10}, MeerLICHT \citep{Bloeman16}, PanSTARRS \citep{Kaiser10}, Searches After Gravitational waves Using ARizona Observatories \citep[SAGUARO;][]{Lundquist19}, the T\'{e}lescope \`{a} Action Rapide pour les Objets Transitoires \citep[TAROT;][]{Boer01}, the Visible and Infrared Survey Telescope for Astronomy \citep[VISTA;][]{Sutherland15}, the VLT Survey Telescope \citep[VST;][]{Capaccioli11} and the Zwicky Transient Facility \citep[ZTF;][]{Bellm19}. No associated transients were identified \citep{Anand20,Antier20,Antier20b,Coughlin20,SaguesCarracedo20}, but constraining limits were placed on a number of milestone events, including S190814bv, the first NSBH merger candidate identified in GW \citep{Dobie19,Gomez19,LSC19,Ackley20,Andreoni20,Vieira20,Watson20}, and several candidate BNS systems \citep{Goldstein19,Hosseinzadeh19,Lundquist19}, including the unusually massive GW190425 \citep{Coughlin19,Hosseinzadeh19,Lundquist19,Abbott20b}.

In this paper, we investigate the 29 follow-up campaigns of GW triggers undertaken by GOTO-4 during the first half of the LVC O3 run (April to October 2019), before GOTO was upgraded to 8 telescopes. The success of GOTO-4 in tiling the LVC error boxes is discussed. While no GW-EM counterpart candidate detections were made, we assess GOTO-4's volumetric coverage in 3D by employing test sources to represent the expected EM signatures accompanying GW events, using the O3a events as a benchmark test sample. The observable horizons are compared to the distance distribution of O3a BNS events. The findings are used to inform future strategy, and highlight areas of focus for future upgrades.

In Section~\ref{sec:data} we further discuss the data acquisition from the LVC archives and the GOTO pipeline. Section~\ref{sec:sources} introduces our test sources, and their application is detailed in Section~\ref{sec:method}. We show our results in Section~\ref{sec:results}, which are discussed in Section~\ref{sec:discussion}. Finally, we present our key conclusions in Section~\ref{sec:conclusions}. We assume a cosmology of \cosmo{} throughout.

\section{Data Sample}\label{sec:data}

\subsection{LVC Superevents}

GW triggers from individual LVC analysis pipelines are aggregated into ``superevents'', which are announced on the Gamma-ray Co-ordinates Network (GCN) and presented on the Gravitational-wave Candidate Event Database (GraceDB\footnote{\label{note1}\url{https://gracedb.ligo.org/}}). Initial position reconstruction of the GW source is performed by the BAYESTAR algorithm \citep{Singer14,Singer15}. The algorithm outputs a Hierarchical Equal Area isoLatitude Pixelization \citep[HEALPix;][]{Gorski05} all-sky map of the posterior probability, as well as the location, scale and normalisation of the conditional distance distribution for each pixel across a grid of millions. At a later time, the BAYESTAR reconstruction may be superseded by a volume reconstruction using LALInference \citep{Aasi13, veitch15}, the Advanced LIGO Bayesian parameter estimation library. Skymaps are made available for download in the form of a Flexible Image Transport System \citep[FITS;][]{Wells81} file. 

Using the HEALPix maps, it is possible to calculate the probability contained in a particular region of sky, or to construct a probability density distribution with distance along a given line of sight \citep[for comprehensive recipes, see][]{Singer16}. Table~\ref{tab:sample} presents the 29 LVC superevents that were followed up by GOTO-4 during the first half of the LVC O3a run (excluding a further 3 triggers that were followed up but later retracted by the LVC). The mean and standard deviation of the all-sky posterior probability distance distribution \citep{Singer16} is shown, along with the classification of each event and the associated false alarm rate (FAR) from GraceDB. We also include the LVC astrophysical classification, which can be BBH (both binary components are constrained to $> 5$~M$_{\odot}$), BNS (both binary components are constrained to $< 3$~M$_{\odot}$), NSBH (one object is constrained to $< 3$~M$_{\odot}$, and the other to $> 5$~M$_{\odot}$) or MassGap (at least one of the binary components is constrained to be between $3$~M$_{\odot}$ and $5$~M$_{\odot}$). The terrestrial (noise) classification category is neglected. In all cases we perform our analysis with the most recently released probability map, and the given distances, classifications and FARs correspond to this map.

\begin{table*}
\begin{tabular}{lcccccccc}
\hline\hline
 & & & \multicolumn{4}{c}{Classification Probability} & & \\
 \cline{4-7}
Event & Distance & $\sigma_{\rm dist}$ & $p_{\rm BBH}$ & $p_{\rm NSBH}$ & $p_{\rm BNS}$ & $p_{\rm MassGap}$ & FAR & Announcement\\
 & Mpc & $\pm$Mpc & \% & \% & \% & \% & (year$^{-1}$) & GCN\\ 
\hline
S190408an & 1473 & 358 & >99 & 0 & 0 & 0 & $8.86\times 10^{-11}$ &  \citetalias{LVCGCNS190408an}\\ 
S190412m & 812 & 194 & 100 & 0 & 0 & 0 & $5.30\times 10^{-20}$ & \citetalias{LVCGCNS190412m}\\ 
S190421ar & 1628 & 535 & 97 & 0 & 0 & 0 & $0.47$ & \citetalias{LVC2GCNS190421ar}\\ 
S190425z & 156 & 41 & 0 & 0 & >99 & 0 & $1.43\times 10^{-5}$ & \citetalias{LVCGCNS190425z}\\
S190426c* & 377 & 100 & 0 & 6 & 24 & 12 & $0.61$  & \citetalias{LVCGCNS190426c}\\
S190510g & 227 & 92 & 0 & 0 & 42 & 0 & $0.28$ & \citetalias{LVCGCNS190510g}\\
S190512at & 1388 & 322 & 99 & 0 & 0 & 0 & $0.06$ & \citetalias{LVCGCNS190512at} \\
S190513bm & 1987 & 501 & 94 &  <1 & 0 & 5 & $1.18\times 10^{-5}$ & \citetalias{LVCGCNS190513bm} \\
S190517h & 2950 & 1038 & 98 & <1 & <1 & 2 & $0.07$ & \citetalias{LVCGCNS190517h} \\
S190519bj & 3154 & 791 & 96 & 0 & 0 & 0 & $0.18$ & \citetalias{LVCGCNS190519bj} \\
S190521g & 3931 & 953 & 97 & 0 & 0 & 0 & $0.12$ & \citetalias{LVCGCNS190521g} \\
S190521r & 1136 & 279 & >99 & 0 & 0 & 0 & $0.01$ & \citetalias{LVCGCNS190521r} \\
S190630ag & 1059 & 307 & 94 & <1 & 0 & 5 & $4.54\times 10^{-6}$  & \citetalias{LVCGCNS190630ag} \\
S190706ai & 5263 & 1402 & 99  & 0 & 0 & 0 & $0.06$ & \citetalias{LVCGCNS190706ai} \\
S190707q & 781 & 211 & >99 & 0 & 0 & 0 & $1.66\times 10^{-4}$ & \citetalias{LVCGCNS190707q}\\
S190718y & 227 & 165 & 0 & 0 & 2 & 0 & $1.15$ &  \citetalias{LVCGCNS190718y} \\
S190720a & 869 & 283 & 99 & 0 & 0 & 0 & $0.12$ & \citetalias{LVCGCNS190720a} \\
S190727h & 2839 & 655 & 92 & <1 & 0 & 3 & $4.35\times 10^{-3}$ & \citetalias{LVCGCNS190727h} \\
S190728q & 874 & 171 & 34 & 14 & 0 & 52 & $7.98\times 10^{-16}$ & \citetalias{LVCGCNS190728q}  \\
S190814bv & 267 & 52 & 0 & >99 & 0 & >1 & $6.40\times 10^{-26}$ & \citetalias{LVCGCNS190814bv}  \\
S190828j & 1803 & 423 & >99 & 0 & 0 & 0 & $2.67\times 10^{-14}$ & \citetalias{LVCGCNS190828j}  \\
S190828l  & 1609 & 426 & 99 & 0 & 0 & 0 & $1.46\times 10^{-3}$ & \citetalias{LVCGCNS190828l}  \\
S190901ap  & 242 & 81 & 0 & 0 & 86 & 14 & $0.22$ & \citetalias{LVCGCNS190901ap}  \\
S190910d & 632 & 186 & 0 & 98 & 0 & 0 & $0.12$ & \citetalias{LVCGCNS190910d}  \\
S190915ak & 1584 & 381 & >99 & 0 & 0 & 0 & $0.03$ & \citetalias{LVCGCNS190915ak}  \\
S190923y & 438 & 113 & 0 & 68 & 0 & 0 & $1.51$ & \citetalias{LVCGCNS190923y}  \\
S190924h & 548 & 112 & 0 & 0 & 0 & 99 & $2.82\times 10^{-11}$ & \citetalias{LVCGCNS190924h}  \\
S190930s & 709 & 191 & 0 & 0 & 0 & 95 & $0.09$ & \citetalias{LVCGCNS190930s} \\
S190930t & 108 & 38 & 0 & 74 & 0 & 0 & $0.49$ & \citetalias{LVCGCNS190930t}  \\
\hline\hline
\end{tabular}

\caption{The sample of LVC superevents that were followed up by GOTO. The distance and $\sigma_{\rm dist}$ columns represent the posterior mean and standard deviation of the distance to the source, marginalised over the whole sky \citep{Singer16}. The classification probabilities and False Alarm Rates (FAR) are taken from GraceDB. Note that ``missing'' probability (i.e. cases where the given probabilities do not sum to 100 per cent) will have been assigned to the ``terrestrial'' (noise) category. *Under the assumption that this source is astrophysical in origin, the classification probability becomes NSBH 12 per cent: MassGap 5 per cent: BNS 3 per cent \citep{GCN24411}.}
\label{tab:sample}
\end{table*}

\subsection{Data Collection with GOTO}

Although we here present an offline analysis of the events, the data collection is driven by a low-latency realtime response.
The LVC all-sky probability map is automatically ingested by the GOTO sentinel \citep{dyer2018telescope}, which analyses the map and produces a schedule of exposures across the observable probability region. These exposures are taken on a fixed grid of tiles on the sky to compare GOTO's new exposure to a reference image of that same patch of the sky. Observations are scheduled as soon as the initial map is available, and updated whenever a map update is released.

GOTO has the capacity to do multi-band photometry, having three colour filters. However, for initial follow-up GOTO uses its wide \textit{L}-band filter. This collects photons between 3750\AA and 7000\AA , roughly equivalent to a combination of GOTO's \textit{B},\textit{G}, and \textit{R} filters and comparable to the combined passband of the SDSS $g$ and $r$ filters.

Each tile is comprised of images from all four telescopes and has a combined field of view of $\approx 19$ square degrees. The default strategy is to visit each tile at least twice per event, where each visit contains a set of 3x60 second exposures, which are conducted back-to-back in the wide \textit{L}-band filter. The three exposures in each set are reduced and then median combined to create the science image. A reference image of the corresponding tile is then subtracted from the science image to identify any new transients. For BNS events, tiles are repeated in order to provide many passes over the skymap over the course of several nights.

\subsection{Image Processing and Data Mining}

Data are processed on a dedicated cluster of machines located in a Warwick server room. The cluster includes high CPU core count processing nodes, storage arrays and database nodes all connected via 10Gbit ethernet. An automated process flow ensures new image frames enter the process queue automatically as they are downloaded from the observatory. For the prototype, a single high-end processing node is sufficient to keep up with processing in realtime (about four 50Mpixel images each minute). The software stack has been developed by the consortium, and performs image level processing, astrometry and photometry calibration, image alignment and subtraction, and source/feature detection. The results are ingested into a PostgreSQL database in realtime to allow vetting of candidates with short latencies (Steeghs et al., in prep).

The initial stages of GOTO's image reduction pipeline cover per-image bias subtraction, dark subtraction, flat-field correction, overscan correction and trimming. Following a source detection pass, an astrometric solution is found using \texttt{astrometry.net} \citep{lang2010astrometry} and photometric zero points are derived through comparison with either APASS V filter \citep{henden2016vizier} or PS1 $g$-filter magnitudes for a large number of field stars. The systematic uncertainty in the zero points varies between Unit Telescopes (UTs) and where a source falls on a given image, but is typically better than $0.15$ mags. A set of exposures (usually 3) are then median-combined, correcting for any astrometric offsets between exposures. For each median stack, a reference image is identified (if available), which then triggers the difference imaging stage, using the \texttt{HOTPANTS} tool \citep{becker2015hotpants}.

Features are detected on the subtracted images and passed to a random forest classifier. This makes use of a number of source attributes such as flux, full-width half maximum, local noise level, etc. in an attempt to filter out artefacts and cosmics. During the prototype phase, GOTO employs a temporary classifier that is similar in structure to that of \citet{Bloom12} and it was trained using an injected source data-set. However, this is an area of active development within the collaboration and will be replaced in the next iteration of the pipeline (Mong et al. in prep; Killestein et al. in prep). Low scoring sources are marked bogus leaving human judgement to vet any remaining high-confidence sources. Features with a reasonable classifier score are then ingested into the GOTO  ``Marshall'', which presents source and contextual information via a browser for human inspection.  This entire process is completed approximately 10-20 minutes after the images are taken in the current prototype pipeline. Human candidate vetting takes place in real-time alongside the follow-up campaign, which typically lasts for several days following a trigger. Any announcements are disseminated via a GCN circulars and/or TNS submissions. For the purpose of this paper, the data are mined after a campaign has been completed using a script which pulls all observations linked to each event. The observations are analysed and their meta-data is taken to assess follow-up performance.

\section{Test Sources}\label{sec:sources}

In order to assess our coverage along the radial distance of the LVC skymap, we must define a test source to recover. For this analysis, we define three physically motivated light curves, as well as a reference source with a constant magnitude. Our physically motivated sources are:
\begin{enumerate}
\item A gamma-ray burst afterglow, viewed along the jet axis (see Section~\ref{sec:grbs}).
\item A gamma-ray burst afterglow viewed off-axis.
\item A Bazin function \citep{bazin2011photometric}, representing a kilonova-like evolution (see Section~\ref{sec:kne}).
\end{enumerate}

These three phenomena are associated with compact object mergers, though none are necessarily expected to accompany the merger of a BBH. However, we note the possible \emph{Fermi}-GBM detection of GW150914 \citep{connaughton2016fermi}, and the subsequent theoretical works that attempt to link (weak) GRBs with BBH mergers, as well as some additional optical phenomena \citep{perna2016short, murase2016ultrafast, de2017electromagnetic, janiuk2017possible, Chang18}. A kilonova-like emission profile remains the only electromagnetic accompaniment to have been detected alongside a GW signal within the wavelength range and timescales covered by GOTO \citep[GW 170817/AT2017gfo;][]{abbott2017multi}, albeit from a neutron star origin. A constant source of $m_L = 19$ is also included in our analysis as a comparison case. This allows us to measure how well GOTO-4 would have performed in retrieving a persistent and reasonably bright new object in a given search field.

Our model light curves are all constructed in the $g$ and $r$ filters. The mean of these two models provides a close match to the GOTO's \textit{L}-band, which is used during GW follow-up. Light curves at a representative 100~Mpc distance are shown in Fig.~\ref{fig:test_sources}.

\subsection{Gamma-Ray Burst Models}\label{sec:grbs}

Our GRB afterglow model light curves are constructed following \citet{Sari1998spectra}. We assume fairly typical physical parameters for sGRBs \citep[e.g.][]{Fong15,gompertz2015broad}. The isotropic equivalent energy in $\gamma$-rays, $E_{\gamma , \rm iso} = 10^{52}$~ergs, the fraction of energy contained in the emitting electrons, $\epsilon_e = 0.1$, and the fraction of energy contained in the magnetic fields, $\epsilon_B = 10^{-2}$. The circumburst environment is assumed to have a constant density with radial distance, with a particle density of $n = 10^{-3}$~cm$^{-3}$. Electrons in the forward shock are assumed to be accelerated into a power law distribution of Lorentz factors with an index of $p = 2.2$. The half-opening angle of the jet is set to $\theta_{\rm o} = 0.087$~rad ($5^{\circ}$). These parameters result in a jet break at $t \approx 1.82$~days \citep[cf.][]{Granot18}.

In the off-axis case, we use the analytical solution from \citet{Granot18} for the jet break time
\begin{equation}
t_{\rm jb} = 0.7(1+z)\bigg(\frac{E_{51}}{n}\bigg)^{1/3} \bigg(\frac{\theta_{\rm o}}{0.1}\bigg)^2 \hbox{days}
\end{equation}
 and peak flux time
 \begin{equation}
    t_{\rm peak}(\theta_{\rm obs}) = \bigg(\frac{\theta_{\rm obs}}{\theta_{\rm o}}\bigg)^2 t_{\rm jb} \hbox{ days,}
 \end{equation}
 where $E_{51} = \theta_{\rm o}^2 E_{\rm iso}/2$ is the beaming-corrected energy in units of $10^{51}$~erg. The flux is assumed to be zero at $t < t_{\rm jb}$ then rises smoothly until its maximum at $t = t_{\rm peak}$. At $t > t_{\rm peak}$ the flux evolves following the standard on-axis evolution. All physical parameters are set to be the same as the on-axis case, including the jet half-opening angle. We assume that the observer is located at an angle of $\theta_{\rm obs} = 0.174$~rad ($10^{\circ}$) from the jet axis. This choice reflects a more optimistic case that provides a (comparatively) bright signal in the absence of a GRB prompt trigger. As the observer moves further from the jet axis, the expected signal becomes fainter, and the light curve peaks later.

\subsection{Kilonova Model}\label{sec:kne}
Our Bazin function model is based on the kilonova AT2017gfo. Following the method of \citet{gompertz2018diversity}, we fit Bazin or exponential functions to the full dataset\footnote{curated on \href{kilonova.space}{kilonova.space} \citep{Guillochon17}} \citep{Andreoni17,Arcavi17,Cowperthwaite17,Diaz17,Drout17,Evans17,Lipunov17,Pian17,Shappee17,Smartt17,Tanvir17,Valenti17,Villar17,Pozanenko18} to obtain the phenomenological evolution of the $g$ (including $g$, F475W and $V$ filters) and $r$ (including $r$, F606W, F625W and $R$ filters) bands. The rise and peak of the $g$ and $r$ bands are unconstrained for AT2017gfo; in both filters the light curve is best fit with an exponential profile that has a decay time of $\tau_{f, g} = 0.94$~days and $\tau_{f, r} = 1.47$~days, and a normalisation of $A_g = 679.5$~$\mu$Jy and $A_r = 631.5$~$\mu$Jy. To avoid over-estimating the flux at early times, we modify both functions from exponential to Bazin by adding a rise time of $\tau_r = 0.1$~days, and $t_0 = t_{\rm max} - \tau_r \ln{}(\tau_f/\tau_r - 1) = 0.3$~days, where $t_{\rm max}$ is the peak emission time. The resulting light curves are consistent with the data, although the rise and peak parameters are not strictly constrained. Our \textit{L}-band magnitudes are derived by assigning an equal weighting to the $g$ and $r$-band models.

\begin{figure}
\includegraphics[width=\columnwidth]{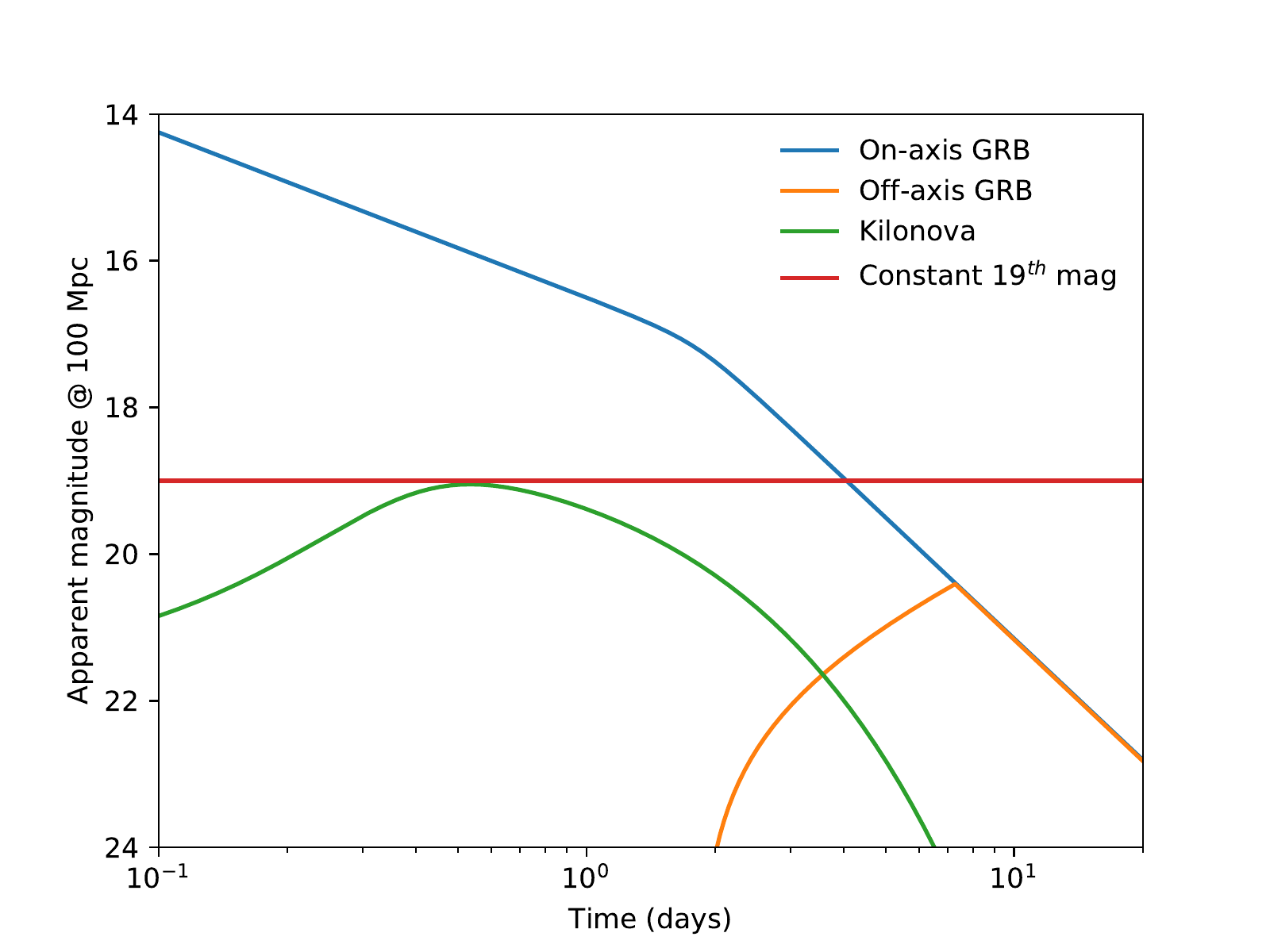}
\caption{Light curves of our test sources at an assumed distance of 100~Mpc.}
\label{fig:test_sources}
\end{figure}

\section{Method}\label{sec:method}

\begin{figure}
\includegraphics[width=\columnwidth]{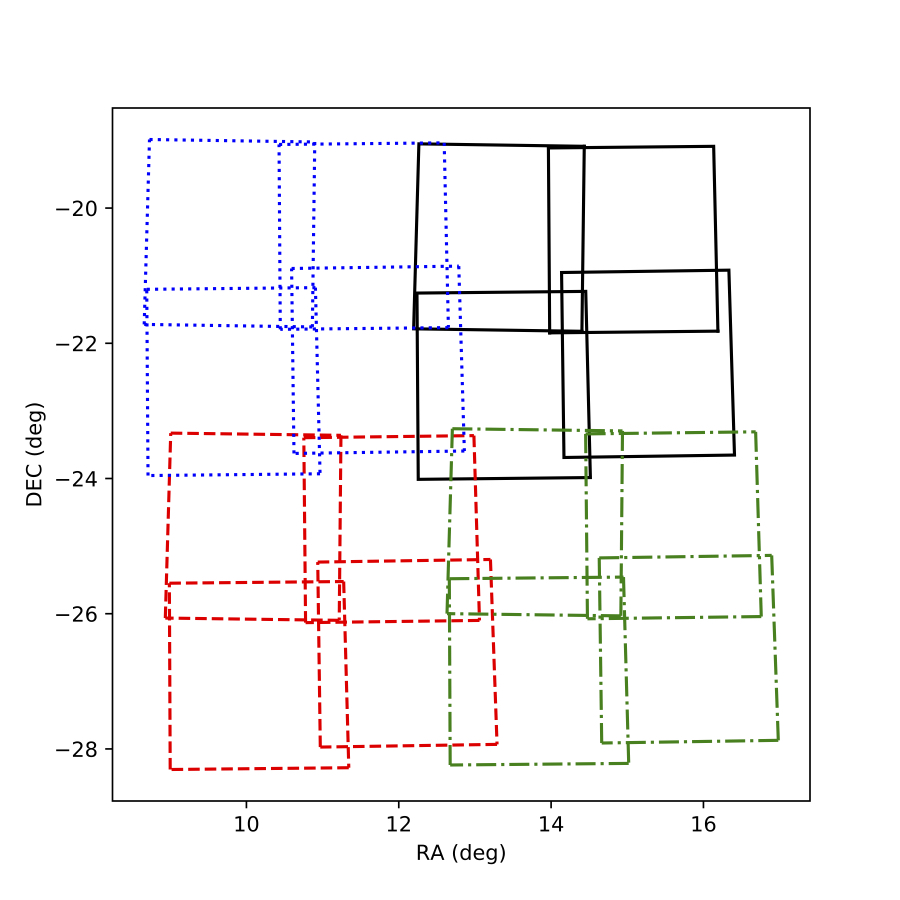}
\caption{Four example GOTO-4 tiles (blue, black, green, red), each of which is comprised of 4 UT snapshots. There is significant overlap in observation both from overlapping tiles and from overlapping UT fields within a given tile. These effects must be accounted for to avoid ``double-counting'' the covered probability.}
\label{fig:overlap}
\end{figure}

We first assess the combined GOTO-4 sky coverage for each event. There are two sources of overlap in the observations: firstly, between individual pointings (tiles) on the sky, and secondly, each tile is the combined footprint of 4 UTs, which themselves overlap one another (see Fig.~\ref{fig:overlap}). While this setup is advantageous because it eliminates the chip gap problem seen in other survey missions (and aids intra-telescope calibration) it makes calculating sky coverage more complex, because simply summing the on-sky footprint of the each image (1/telecope) results in a lot of double (or more) counting. Fortunately, a natural solution to double counting coverage is available in the form of the pixel grid in the LVC HEALPix maps. The Healpy\footnote{\url{https://healpy.readthedocs.io/en/stable/index.html}} {\sc query\_polygon} routine returns the indices of pixels contained within a user-defined polygon, meaning that we can generate a pixel coverage list by entering the coordinates of the corners of each individual UT snapshot. GOTO-4 sky coverage is then calculated by summing the area of all unique pixel instances in the list. Similarly, the total probability covered is calculated by summing the probability contained within every unique pixel covered. We calculate coverage with the ``inclusive'' keyword set to False when running {\sc query\_polygon}, meaning that only pixels with centres that lie within our observed area are counted. The typical pixel size of the HEALPix maps is $\approx 10$ square arc minutes; much smaller than the total areas counted.

\subsection{Volumetric Coverage}\label{sec:source_coverage}

To assess our volumetric coverage, we compare our test source models (Section~\ref{sec:sources}) at each observing epoch to the limiting magnitude of the science image in order to assess from how far away they could be detected. Model magnitudes are corrected for Galactic extinction using an all-sky foreground reddening, $(E(B-V))$, HEALPix map based on the results of \citet{Schlegel98}, which is available for download from the Goddard Space Flight Center website\footnote{\href{lambda.gsfc.nasa.gov/product/foreground/fg\_sfd\_get.cfm}{lambda.gsfc.nasa.gov/product/foreground/fg\_sfd\_get.cfm}}. Extinction corrections are applied based on which pixel of the \emph{reddening} map each LVC \emph{probability} map pixel falls into. The two maps typically have comparable resolution (within a factor of two), but variability in the dust maps on scales of less than $\approx 10$ arc minutes$^2$ is lost due to being averaged over the pixel. \citet{Schlegel98} values of $E(B-V)$ are converted to the more recent findings of \citet{Schlafly11} using $E(B-V)_{\rm SF11} = 0.86\times E(B-V)_{\rm SFD98}$ \citep{Schlafly10,Schlafly11}.

\begin{figure}
    \centering
    \includegraphics[width=\linewidth]{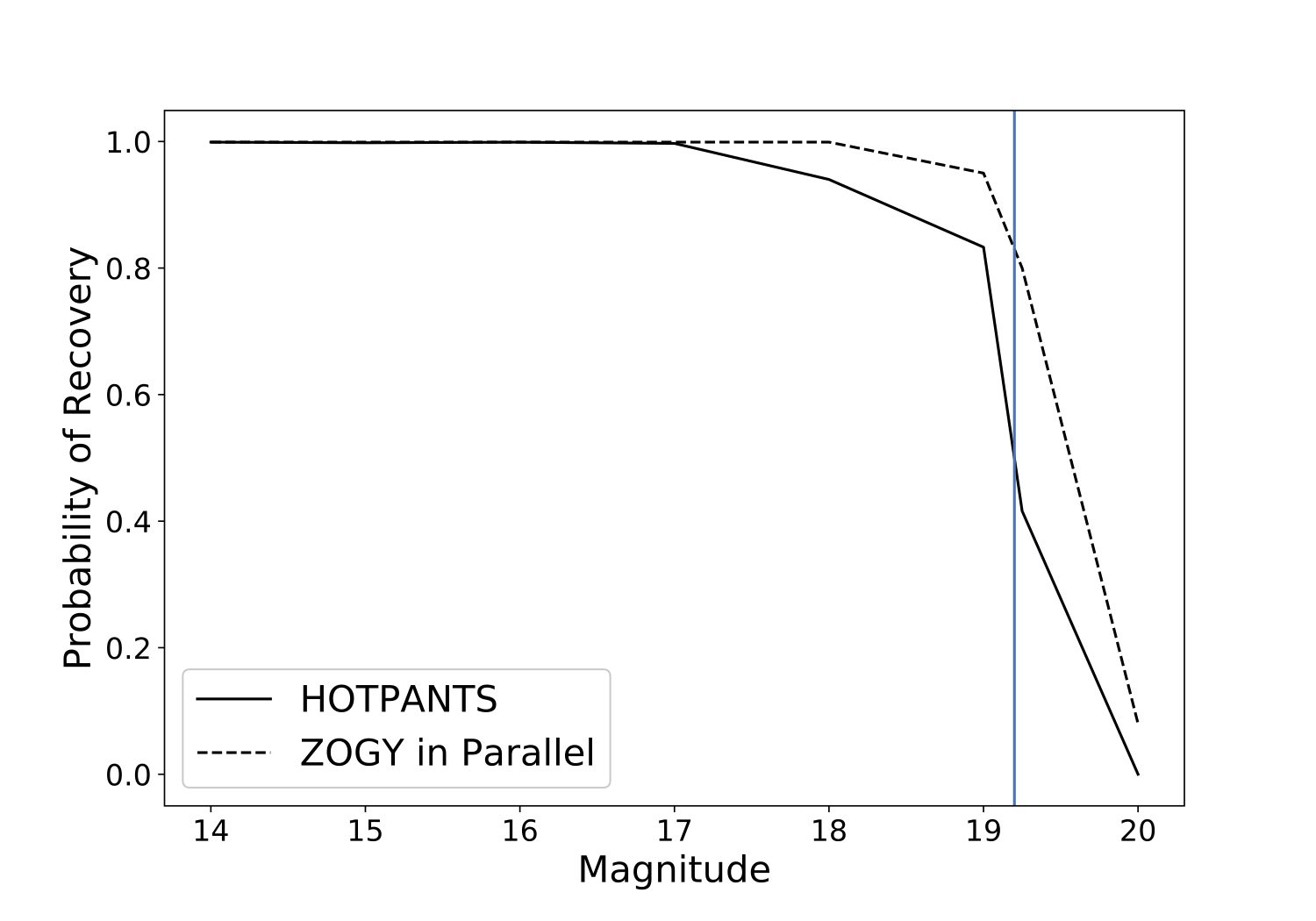}
    \caption{The fraction of artificially injected test sources recovered vs magnitude, down to the average $m_{\rm lim}$ (vertical blue line). In the optimal case nearly 80 per cent of sources are recovered at this limit (horizontal line). The steep turnover indicates that a hard cutoff at $m_{\rm lim}$ is a suitable working approximation for the observable horizon.}
    \label{fig:RECOVERY}
\end{figure}

Next, we calculate the horizon out to which our extinction-corrected test-source magnitude could be detected by GOTO-4 for each observation. The horizon is defined as the distance from which the extinction-corrected test-source magnitude would be equal to the 5$\sigma$ upper limit of a given image, $m_{\rm lim}$. This limit is based off the science frame. This determines the optimal depth the transient search can achieve from image subtraction (assuming the reference image can see deeper). $m_{\rm lim}$ is stored in the FITS header of the GOTO-4 observation, and is defined using the pipeline zero-point derived limiting magnitude. Where there are multiple visits to the same LVC probability map pixel, we take the visit with the farthest horizon, and discard the duplicates. Our definition of the horizon implicitly treats $m_{\rm lim}$ as a ``hard'' cutoff, where everything brighter is detected and everything fainter is missed. In practice, $m_{\rm lim}$ represents the average magnitude limit for a given image, which means that in some cases brighter objects would be missed within the horizon, and fainter objects could be detected beyond it (e.g. due to background noise fluctuations, proximity to bright stars, etc). However, the recovery curve, while not a box function, is found to be very steep (Fig. \ref{fig:RECOVERY}), making our hard limit assumption a useful working approximation.  In brief, the recovery curve was built by injecting a large number of mock transients into the images, using a large sample of images from the O3a observations and covering a wide range of magnitudes. These fields are then image subtracted, using two different methods, and an automatic source extraction test is done using \texttt{SExtractor}. If a source is found within 2 pixels of its injected location it is counted as recovered. Full details of the process are available in Cutter et al. (in prep). The output shown here is expected of a typical GOTO exposure. For our analysis we show both the HOTPANTS curve that is close to the prototype pipeline performance, as well as the ZOGY in Parallel implementation that is under development. 

This again supports our working assumption in terms of using our O3 campaigns as guidance towards future strategy, and performance will be further improved. We also note that variations in $m_{\rm lim}$ across the images are significantly larger than this particular effect, and is the dominant contribution in terms of the overall coverage achieved.

Finally, LVC probability map pixels are sorted into groups of equal observable horizon, where their probability density functions are summed \citep[cf.][]{Singer16}, and the combined probability density function of each group is integrated out to their shared horizon. Our full volumetric probability coverage is then the sum of all of the groups.

\subsection{Coverage of an AT2017gfo-like Event}\label{sec:2017gfo}

While it is important to assess our coverage of all existing GW events, the main targets of EM follow-up facilities like GOTO are BNS mergers that are (relatively) nearby, since these events are the most likely to produce a detectable signature in the form of a KN. AT2017gfo remains the only KN detected alongside a BNS merger that was confirmed by GW detections, and hence is the gold standard. We therefore also assess our coverage and follow-up strategy by calculating the maximum observable horizon of an AT2017gfo-like event in each LVC probability map, and how much of the 2D probability coverage GOTO obtained can be recovered when increasing the assumed distance to the test source. Note that this test is done \textit{in addition} to the test described in Section~\ref{sec:source_coverage}, which was performed on all four test models including our KN analogue.

The method is largely the same as in Section~\ref{sec:source_coverage}, except that the test source is always the composite $g$- and $r$-band Bazin function fit to AT2017gfo. Instead of constructing and integrating the probability density functions from the LVC skymap, we calculate the expected extinction-corrected magnitude of AT2017gfo in each HEALPix pixel for distance increments out to 500~Mpc. These magnitudes are compared to $m_{\rm lim}$, and the column in each line of sight is considered to be covered for the given distance where the magnitude is brighter.

\begin{table*}
\tabcolsep=0.15cm
\begin{tabular}{cccccccccccccccc}
\hline\hline

 & \multicolumn{2}{c}{Response Time} & & \multicolumn{3}{c}{2D Coverage} & & \multicolumn{4}{c}{3D Coverage} & & \multicolumn{3}{c}{KN Range} \\
\cline{2-3} \cline{5-7} \cline{9-12} \cline{14-16}
Event & $\delta t_{\rm trig}$ & $\delta t_{\rm alert}$ & & Area & $pA$ & $pA_{\rm vis}$ & & $pV_{\rm bazin}$ & $pV_{\rm GRB}$ & $pV_{\rm off-axis}$ & $pV_{\rm c19}$ & & $D_{90}$ & D$_{50}$ & D$_{0}$ \\
 & (hours) & (hours) & & (deg$^2$) & (\%) & (\%) & & (\%) & (\%) & (\%) & (\%) & & (Mpc) & (Mpc) & (Mpc) \\
\hline
S190408an$^\dagger$ & 11.4 & 10.8 & & 156.1 & 20.2 & 23.8 & & $1.20\times 10^{-5}$ & $1.47\times 10^{-2}$ & $2.82\times 10^{-7}$ & $3.22\times 10^{-5}$ & & 31 & 70 & 135 \\
S190412m$^\dagger$ & 15.0 & 14.0 & & 295.2 & 94.4 & 94.7 & & $8.68\times 10^{-3}$ & $3.48$ & $0$ & $1.07\times 10^{-2}$ & & 107 & 117 & 151 \\
S190421ar & 48.3 & 29.1 & & 114.3 & 8.88 & 36.6 & & $4.92\times 10^{-5}$ & $3.97\times 10^{-3}$ & $3.89\times 10^{-7}$ & $3.49\times 10^{-4}$ & & 57 & 61 & 66 \\
S190425z & 12.4 & 9.50 & & 2667.1 &  22.0 & 38.1 & & $5.90$ & $20.6$ & $2.57\times 10^{-3}$ & $8.10$ & & 46 & 134 & 227 \\
S190426c & 5.30 & 5.00 & & 772.7 &  54.1 & 70.2 & & $1.10\times 10^{-2}$ & $8.98$ & $0$ & $1.42\times 10^{-2}$ & & 4 & 44 & 136 \\
S190510g & 1.42 & 0.40 & & 116.1 &  0.21 & 0.55 & & $2.06\times 10^{-3}$ & $0.21$ & $0$ & $3.60\times 10^{-2}$ & & 48 & 55 & 57 \\
S190512at & 2.78 & 2.50 & & 315.1 &  87.1 & 92.4 & & $8.52\times 10^{-5}$ & $0.37$ & $0$ & $1.26\times 10^{-4}$ & & 22 & 60 & 154 \\
S190513bm$^\dagger$ & 0.55 & 0.05 & & 116.2 & 28.5 & 76.3 & & $1.35\times 10^{-5}$ & $0.59$ & $0$ & $2.51\times 10^{-5}$ & & 56 & 83 & 120 \\
S190517h$^\dagger$ & 15.9 & 15.2 & & 112.7 & 14.8 & 51.6 & & $1.40\times 10^{-6}$ & $1.25\times 10^{-4}$ & $0$ & $1.62\times 10^{-6}$ & & 49 & 67 & 84 \\
S190519bj$^\dagger$ & 5.35 & 4.35 & & 664.8 & 84.7 & 85.3 & & $2.41\times 10^{-6}$ & $9.55\times 10^{-4}$ & $0$ & $3.64\times 10^{-6}$ & & 43 & 69 & 161 \\
S190521g & 0.13 & 0.05 & & 393.2 & 43.7 & 86.7 & & $8.30\times 10^{-6}$ & $7.57\times 10^{-2}$ & $0$ & $1.11\times 10^{-5}$ & & 94 & 107 & 126 \\
S190521r$^\dagger$ & 15.2 & 15.1 & & 720.7 & 91.9 &  92.9  & & $3.85\times 10^{-6}$ & $1.17\times 10^{-3}$ & $0$ & $7.32\times 10^{-6}$ & & 9 & 51 & 93 \\
S190630ag & 2.40 & 2.40 & & 1170.3 & 60.9 & 79.5 & & $1.33\times 10^{-3}$ & $19.0$ & $1.66\times 10^{-7}$ & $3.09\times 10^{-3}$ & & 71 & 112 & 150 \\
S190706ai & 0.33 & 0.03 & & 543.9 & 36.7 & 48.5 & & $8.03\times 10^{-6}$ & $1.07$ & $1.67\times 10^{-8}$ & $2.86\times 10^{-5}$ & & 55 & 94 & 168 \\
S190707q & 12.4 & 11.7 & & 722.9 & 34.4 & 59.3 & & $2.06\times 10^{-5}$ & $2.77\times 10^{-2}$ & $0$ & $2.54\times 10^{-5}$ & & 18 & 53 & 122 \\
S190718y$^\dagger$ & 6.58 & 6.10 & & 242.5 & 61.2 & 72.9 & & $1.12$ & $28.9$ & $1.54\times 10^{-2}$ & $2.45$ & & 10 & 27 & 90 \\
S190720a & 0.08 & 0.04 & & 1358.3 & 62.1 & 73.3 & & $1.89\times 10^{-4}$ & $9.51$ & $7.67\times 10^{-7}$ & $5.45\times 10^{-4}$ & & 42 & 54 & 163 \\
S190727h & 15.0 & 14.9 & & 714.7 & 42.3 & 93.5 & & $5.72\times 10^{-7}$ & $6.03\times 10^{-5}$ & $0$ & $1.43\times 10^{-6}$ & & 52 & 66 & 140 \\
S190728q & 14.8 & 14.5 & & 146.9 & 89.5 & 94.0 & & $5.55\times 10^{-4}$ & $1.03$ & $0$ & $8.62\times 10^{-4}$ & & 114 & 124 & 139 \\
S190814bv & 1.83 & 1.50 & & 717.9 & 94.1 & 99.1 & & $1.23\times 10^{-2}$ & $89.6$ & $2.33\times 10^{-6}$ & $2.12\times 10^{-2}$ & & 55 & 61 & 81 \\
S190828j & 16.1 & 15.8 & & 442.2 & 9.11 & 81.6 & & $1.01\times 10^{-5}$ & $2.30\times 10^{-3}$ & $6.45\times 10^{-8}$ & $1.27\times 10^{-5}$ & & 34 & 105 & 149 \\
S190828l & 16.9 & 16.5 & & 453.6 & 1.94 & 50.5 & & $5.60\times 10^{-5}$ & $9.20\times 10^{-3}$ & $4.66\times 10^{-7}$ & $7.34\times 10^{-5}$ & & 127 & 138 & 154 \\
S190901ap & 0.12 & 0.04 & & 2523.5 & 38.3 & 45.3 & & $0.34$ & $30.2$ & $8.40\times 10^{-4}$ & $1.16$ & & 62 & 88 & 144 \\
S190910d & 0.13 & 0.03 & & 1675.0 & 41.2 & 85.1 & & $5.43\times 10^{-3}$ & $17.6$ & $0$ & $1.87\times 10^{-2}$ & & 28 & 69 & 148 \\
S190915ak & 29.9 & 29.8 & & 18.2 & 0.08 & 0.08 & & $3.63\times 10^{-11}$ & $2.39\times 10^{-9}$ & $0$ & $8.42\times 10^{-11}$ & & 10 & 10 & 15 \\
S190923y$^\dagger$& 13.8 & 13.7 & & 723.7 & 39.4 & 59.7 & & $1.91\times 10^{-2}$ & $8.95$ & $0$ & $2.29\times 10^{-2}$ & & 46 & 95 & 120 \\
S190924h & 2.97 & 2.90 & & 281.3 & 70.2 & 73.1 & & $4.52\times 10^{-5}$ & $26.4$ & $5.05\times 10^{-8}$ & $3.59\times 10^{-4}$ & & 61 & 75 & 101 \\
S190930s & 6.28 & 6.20 & & 2139.9 & 92.2 & 92.2 & & $2.20\times 10^{-3}$ & $14.2$ & $1.06\times 10^{-6}$ & $4.48\times 10^{-3}$ & & 13 & 89 & 142 \\
S190930t$^\dagger$ & 12.8 & 12.7 & & 918.2 & 6.84 & 9.91 & & $1.24$ & $6.55$ & $1.06\times 10^{-3}$ & $2.01$ & & 48 & 109 & 130 \\
\hline
Mean & 9.90 & 8.79 & & 732.3 & 45.3 & 64.4 & & $0.30$ & $9.91$ & $6.87\times 10^{-4}$ & $0.48$ & & 48 & 79 & 126 \\
Median & 6.58 & 6.20 & & 543.9 & 41.2 & 73.1 & & $8.52\times 10^{-5}$ & 1.03 & 0 & $3.59\times 10^{-4}$ & & 48 & 70 & 136 \\
\hline\hline
\end{tabular}
\caption{GOTO-4 coverage of the LVC probability maps. $\delta t_{\rm trig}$ is the time between the GW trigger and the first GOTO-4 observation. $\delta t_{\rm alert}$ is the time between receiving the LVC preliminary notification and the first GOTO-4 observation. $pA$ is the percentage of the total probability that was tiled by GOTO-4. $pA_{\rm vis}$ represents the percentage of the total probability that was visible to GOTO-4 from its site in La Palma, accounting for Sun constraints and altitude limits. The 3D Coverage columns indicate the volumetric coverage for each of the test sources defined in Section~\ref{sec:sources}. The KN Range columns indicate the horizon out to which 90, 50 and zero per cent of the 2D probability coverage is retained in a search for an AT2017gfo-like event.$\dagger$ denotes that a BAYESTAR map was used; no LALInference map was available.}
\label{tab:results}
\end{table*}

\begin{figure*}
    \centering
    \includegraphics[width=\linewidth]{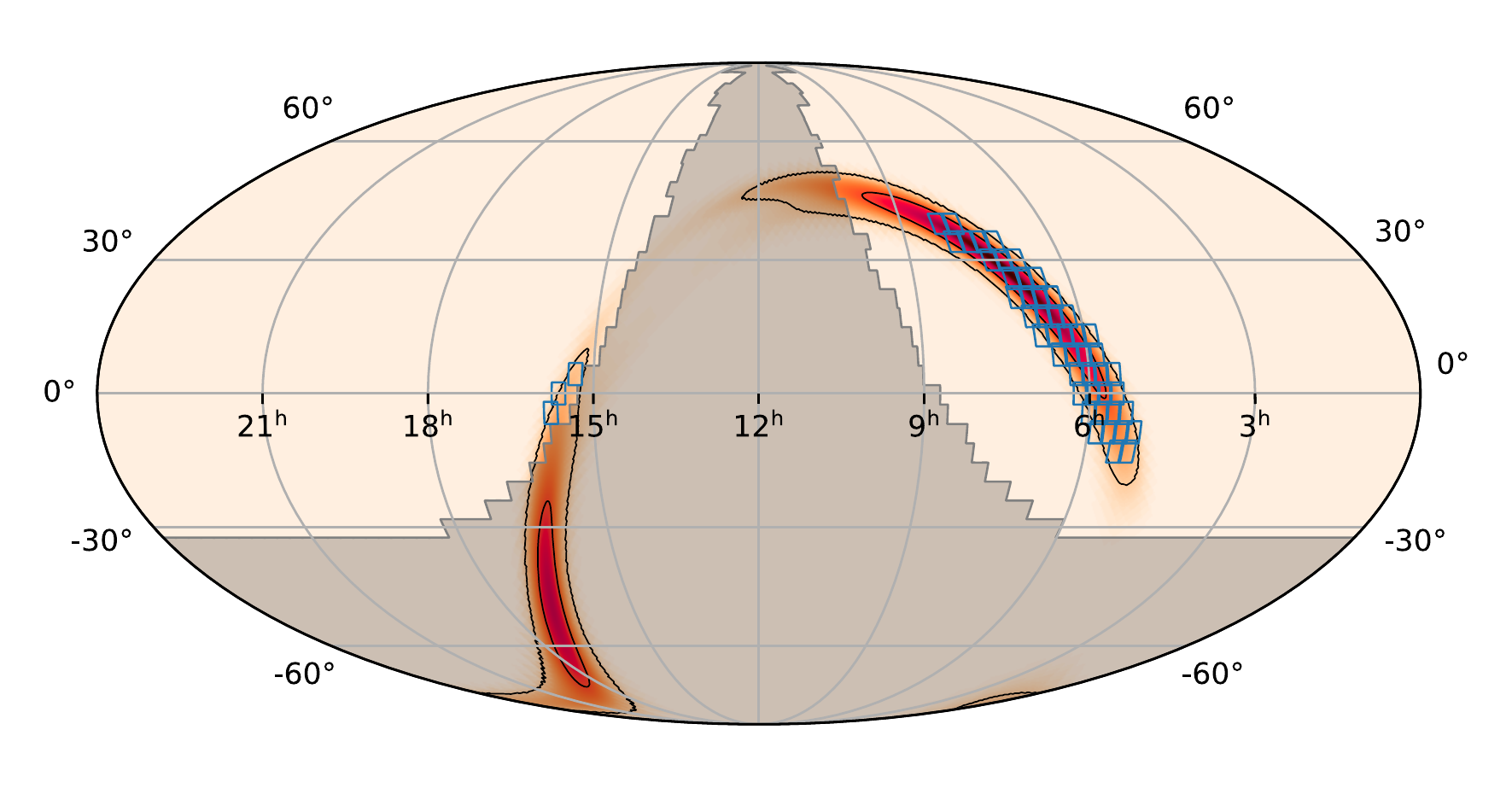}
    \caption{GOTO-4 follow-up of S190923y (\citetalias{LVCGCNS190923y}). Each blue box represents one GOTO-4 tile, with the observing strategy prioritising the highest probability tiles first (darker shaded regions are higher probability). The Area Covered and $pA$ values in Table~\ref{tab:results} comprise the sum of the physical area and probability contained within these tiles respectively, after accounting for overlap. The shaded area indicates unobservable regions, either from altitude limits or Sun constraints. $pA_{\rm vis}$ (Table~\ref{tab:results}) indicates the total fraction of probability that is observable (i.e. that lies within the non-grey region). Axes are in RA/Dec.}
    \label{fig:S190923y}
\end{figure*}

\section{Results}\label{sec:results}

Results of the 29 GOTO-4 follow-up campaigns from the first half of the LVC O3 run are shown in Table~\ref{tab:results}. On average, GOTO-4 began observations $8.79$ hours after receiving an LVC preliminary alert ($9.90$ hours after trigger), and tiled $732.3$ square degrees of the sky per LVC superevent, encompassing $45.3$ per cent of the localisation probability. When observations were unconstrained at the time of the alert (i.e. night time, with the field above the altitude limit), GOTO-4 began observations less than a minute after receiving the LVC preliminary notice. As an example, Fig.~\ref{fig:S190923y} illustrates our follow-up of S190923y, which represents the `mean' campaign given that we covered $723.7$ square degrees and $39.4$ per cent of the probability in this case (see Table~\ref{tab:results}). We also calculate the total probability available to GOTO-4 from its site on La Palma for each event, i.e. the probability that lies within the region of the sky above the telescope's minimum altitude constraint (30 degrees) and away from the Sun. These are shown as pA$_{\rm vis}$ in Table~\ref{tab:results}. On average, $64.4$ per cent of the total probability was observable for a given event over a 2 to 3 night period, meaning that GOTO-4 tiled $\approx 70.3$ per cent of the \emph{available} probability per event on average.

Notably, GOTO-4 tiled more than 500 square degrees on more than half of its campaigns (15), and more than 1000 square degrees on 6 different occasions. This achievement clearly highlights the value of rapid-response, wide-field instruments when searching for poorly-localised transient events. GOTO-4's volumetric coverage is also presented in Table~\ref{tab:results}, expressed as the percentage of the three dimensional probability map probed for each test source. The recovery performance is predictably diminished; EM follow-up facilities are not designed to search for predominantly BBH mergers at a mean distance of $1.3$~Gpc. However, GOTO-4 was able to cover almost 10 per cent of the total probability \emph{volume} on average for an on-axis GRB model. We can rule out our test case on-axis GRB in the observed area in three events (S190510g, S190814bv and S190930t) to better than $p < 0.05$. In order to place an approximate limit on emission from a BBH merger, we take the mean of the least constraining observation within one day after trigger across every BBH follow-up campaign. This is a $5\sigma$ upper limit of $m_L \gtrsim 19$ on emission from a BBH merger inside the GOTO-4 field of view and during the GOTO observing epochs at times of less than 1 day after GW trigger.

Such wide searches do uncover a significant number of candidate detections. Following the automated classifier mentioned previously, and an automatic check against known MPC objects (which otherwise would dominate), we are typically presented with on average one high confidence transient candidate per 15 deg$^2$ during the O3a searches. Contextual information from external catalogs is used to sort these and flag interesting candidates in the context of the GW search. False positives include un-catalogued variable stars and flare stars, which were ruled out as counterparts via comparison with PS1 images.
Many other candidates were disregarded as potential counterparts using the on-sky proximity and redshift of any nearby galaxies through the GLADE catalogue \citep{dalya2018glade}, or via a pre-trigger reported detection either in our own detections tables, or already reported by others as a known transient in the Transient Name Server (TNS). 

Some of the GOTO photometry of unassociated and uncatalogued transients were reported to the TNS where they were high confidence. However, since O3a was very much a test run for the project and its transient classification stack, we relied on several layers of human confirmation and also required multiple detections. Furthermore, automated TNS submission was not in place and generally we did not report photometry if other projects had already done so, unless it turned out to be a true object of interest. We intend to speed-up this process and improve our classifier robustness going forward so that we can increase our chances of being the first to report.

The expected number of events of each type in our sample are presented in Table~\ref{tab:expectations}. We present the results for the localisation map as a whole, for the area tiled by GOTO-4, and for those events closer than 250~Mpc, which are our main targets of interest with GOTO. For off-axis GRBs, KNe and our constant 19$^{\rm th}$ magnitude test sources, the expected rates are essentially unchanged when limiting our results to those events within 250~Mpc. This fact indicates that distant (> 250~Mpc) events do not contribute to the probability of detecting sources of this type. In contrast, the chances of detecting an on-axis GRB is greatly diminished when limiting the search volume; their much greater luminosities mean that they can be detected from much further away.

The GW population is dominated by distant BBH mergers, but for nearby ($< 250$~Mpc) triggers where the majority of the probability map can be observed and tiled in a two night campaign, GOTO-4 can expect to detect a KN similar to AT2017gfo in fewer than 5 campaigns if one is present in each skymap. This is due to the still relatively poor localisation performance of the current GW detector network, coupled with the relatively limited coverage of the GOTO-4 prototype. This will improve substantially in the future \citep{Abbott20}, as further discussed below.

\begin{table}
    \centering
    \begin{tabular}{rccccc}
    \hline\hline
     & Events & GRB & Off-axis GRB & KN & 19$^{\rm th}$ mag \\
    \hline
        Whole map & 29 & $2.87$ & $10^{-4}$ & $0.09$ & $0.14$ \\
        Obs & 29 & $7.04$ & $3\times 10^{-4}$ & $0.49$ & $0.90$ \\
        Nearby & 5 & $0.86$ & $10^{-4}$ & $0.09$ & $0.14$ \\
        Nearby, obs & 5 & $4.15$ & $3\times 10^{-4}$ & $0.49$ & $0.90$ \\
        BBH & 17 & $0.34$ & $10^{-8}$ & $10^{-4}$ & $2\times 10^{-4}$ \\
        BNS & 5 & $0.89$ & $10^{-4}$ & $0.07$ & $0.12$ \\
        All non-BBH & 12 & $2.53$ & $10^{-4}$ & $0.09$ & $0.14$ \\
        \hline\hline
    \end{tabular}
    \caption{The number of detections of each type of signal we would expect to have seen in our sample and highlighted sub-samples if all LVC superevents contained our test sources. ``Whole map'' assumes that the test source could be anywhere in the map, whether the region was observed or not. ``Obs'' assumes that the test source occurred within the region covered by GOTO-4. ``Nearby'' refers to events that occured within 250~Mpc. ``BBH'', ``BNS'' and ``All non-BBH'' are given for the ``Whole map'' scenario, but are limited to events where the given classification was assigned the highest probability by the LVC, excluding terrestrial.}
    \label{tab:expectations}
\end{table}

\begin{figure}
\includegraphics[width=\columnwidth]{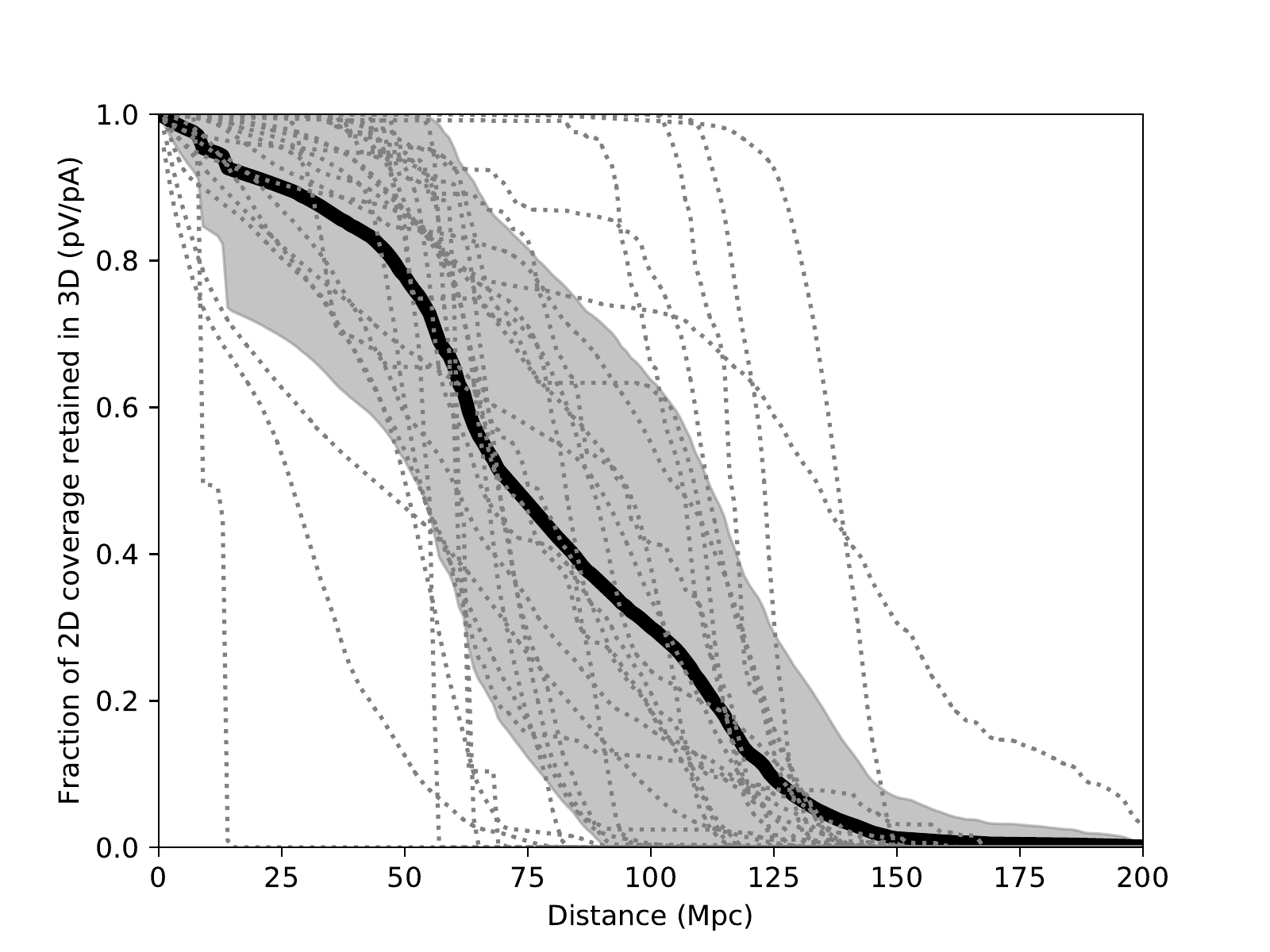}
\caption{The fraction of the 2D probability coverage that is retained with increasing distance. Each dotted line represents an LVC follow-up campaign with GOTO-4. The thick black line is the mean of the 29 events, and the grey area is the $1\sigma$ standard deviation.}
\label{fig:pVpA}
\end{figure}

Table~\ref{tab:results} includes the horizons out to which we retain 90, 50 and 0 per cent of our 2D coverage for an AT2017gfo analogue, following the method in Section~\ref{sec:2017gfo}. For example, if GOTO covered 100 per cent of the probability in the LVC skymap in 2D, $D_{90}$ represents the distance (in Mpc) out to which we probe deep enough to still cover 90 per cent of the probability -- where the lost 10 per cent lies along sight lines that were not probed with sufficient depth to recover the transient. $D_{50}$ indicates that 50 per cent of the observed probability is retained, and $D_{0}$ represents the distance at which no GOTO-4 observation was sufficiently deep to recover an AT2017gfo-like KN. This `completeness' versus distance for each event is also shown in Fig.~\ref{fig:pVpA}, along with the mean of all events. These results show that with reasonable observing conditions, GOTO-4 can detect an AT2017gfo-like event out beyond 100~Mpc, and is capable of achieving 200~Mpc in a favourable line of sight. Of further encouragement is the duplication factor; on average, GOTO-4 observed a given LVC skymap pixel $4.8$ times during a campaign (due to a combination of repeat visits and tile/UT overlap), meaning that there is a great deal of scope to improve the observable horizon through image stacking. In some cases, however, the observable horizon is held back by poor observing conditions. These conditions include clouds, high airmass, a bright moon, or even an unfavourable Galactic pointing resulting in high extinction along the line of sight.

\section{Discussion}\label{sec:discussion}

\subsection{General Constraints}
We find that limits on EM emission from the O3a population of BBH mergers using GOTO-4 are currently unconstraining. Under the assumption that every event does contain one of our test sources (an assumption that is extremely unlikely to be true in practice), we would require more than $10^5$ BBH events to place any meaningful limits with our O3a configuration. However, with information on the inclination of the system, we may be able to constrain BBH mergers as GRB progenitors after fewer than 50 events. GOTO-4 covered almost 10 per cent of the LVC probability volume on average for an on-axis GRB model during O3a, and hence we would expect to find an on-axis GRB in $\approx$ one in ten events if one is present in each case. These constraints will tighten as coverage is significantly improved going forward, thanks to the scaling up of GOTO and the improved localisation precision from the LVC. More generally, GOTO-4 places a mean $5\sigma$ limit of $m_L \gtrsim 19$ in the observed area within one observer frame day from trigger for all BBH events (where observations were taken inside this window). This figure comes from taking the mean value of the least constraining tile from each campaign. However, it does not include foreground extinction, and the observations do not cover the full 24 hour period.

Our analysis suggests that the current GOTO-4 follow-up strategy for GW triggers is not suited for placing constraints on associated off-axis GRBs. This limitation arises for two reasons: the first is that they are faint compared to the other expected transients (see Fig.~\ref{fig:test_sources}) and the second is that the GOTO follow-up strategy focuses on the first two or three nights after the trigger (as is appropriate for KNe), whereas off-axis GRBs will peak several days later than this. The time of observation is the more impactful of the two factors, and we still expect to find untriggered off-axis GRBs in the survey data, because their intrinsic brightness lies well within GOTO's capability. In the future, GOTO will be able to go deeper faster as more telescopes are deployed, and the LVC localisations will improve, leaving smaller error regions to search. Our findings here suggest that the GOTO GW follow-up strategy may benefit from returning to a candidate field after $\sim 5$ -- $10$~days post-trigger, in order to better constrain the possibility of an off-axis GRB. Such discoveries were indeed made during O3 \citep{kool19,perley19}, highlighting the scientific potential for revising the strategy.

Our KN model is based on AT2017gfo, a confirmed BNS merger \citep{Abbott17b}. NSBH mergers are also expected to produce KNe, and their emission profiles may be quite different. If the NS plunges directly into the BH without material being disrupted \citep[the fate of the NS depends primarily on the binary mass ratio and the BH spin;][]{Shibata11,Kawaguchi15,Kyutoku15,Foucart18}, then no EM transient is expected at all. The LVC analysis suggested that this was the case for all of the NSBH merger candidates during O3a \citep[S190814bv, S190910d, S190923y and S190930t;][]{LVCGCNS190814bv,LVCGCNS190910d,LVCGCNS190923y,LVCGCNS190930t}. If tidal stripping occurs during inspiral, then material may remain outside of the BH event horizon, and a KN can be produced. In the optical frequencies at which GOTO-4 observes, this KN may in fact be fainter than an AT2017gfo analogue, unless a significant ($\geq 0.02$M$_{\odot}$) amount of matter is ejected \citep[e.g.][]{Kawaguchi20,Zhu20}. However, for a larger ejecta mass, this signal may in fact be brighter instead. A KN from an NSBH merger may therefore be brighter than AT2017gfo, fainter, or absent entirely. Like with BNS mergers, we emphasise that a significant diversity of light curves are likely to be produced in nature, and that our use of AT2017gfo (as the best-studied and most concrete example) is as a known representative.

\subsection{The Binary Neutron Star Population}
When focusing on nearby (< 250~Mpc) events, or those classified as a BNS merger (where BNS is the most probable classification, excluding terrestrial), the prospects of a detection are naturally greatly improved. Table~\ref{tab:results} shows that the average GOTO-4 follow-up campaign probes sufficiently deeply to detect AT2017gfo \citep[$D_L = 41 \pm 3.1$~Mpc;][]{Hjorth17} because the mean $D_{90} = 48$~Mpc. This means that 90 per cent of observed sight lines, by probability contained, are probed deeply enough to detect AT2017gfo at a distance of 48~Mpc. Similarly, per Table~\ref{tab:results}, the average response time of $9.90$ hours is fast enough to catch the peak of the emission, and typically $64.4$ per cent of the available probability is covered. In favourable conditions the telescope can detect a similar event out to $\sim 200$~Mpc. Notably, Table~\ref{tab:expectations} shows that the chances of detecting a KN is a strong function of the 2D coverage, given the large jump in probability when we calculate our expectations for the surveyed area. 
This fact will be further exploited by the expansion of the GOTO network, both in terms of an increased number of telescopes as well as a 2nd node in the southern hemisphere. When fully deployed, this amounts to 2x16 scopes, a factor of 8 more than the GOTO-4 prototype presented here. This will allow faster tiling and more comprehensive and deeper coverage.

Table~\ref{tab:expectations} indicates that in its current configuration, with no stacking, GOTO-4 can expect to find one AT2017gfo-like KN per few tens of events observed within 250~Mpc. This relatively low estimated discovery rate is primarily due to the O3a BNS sample being more distant on average than pre-O3 expectations, where the `most likely' estimate of \citet{Abadie10} suggested $\sim 4$ BNS mergers per year within 100~Mpc (with expectations ranging from $0.04$ to $42$ events). More recent rate estimations following the LVC O1 and O2 runs predicted $0.5$ -- $16$ BNS mergers per year \citep{Abbott19}, and the most recent estimate from the LVC, which includes GW190425 \citep[here S190425z]{LIGO20}, constrained the rate of BNS mergers within 100~Mpc to be between $1$ and $12$ events per year.

A na\"{i}ve estimate, based on the 5 candidate BNS mergers during the first half of O3 ($\sim 6$ months) and using the most distant detection (377~Mpc; S190426c) to define the radius of the observed volume, implies a rate of $\sim 4.5 \times 10^{-8}$~Mpc$^{-3}$yr$^{-1}$. This indicates that based on the observed population of candidate BNS mergers during the first half of O3, only $0.1$ BNS merger events within 100~Mpc are expected; towards the lower end of the \citet{Abadie10,Abbott19} and \citet{LIGO20} rate estimates. However, our estimate assumes that the GW detector network was always on during O3a, and does not account for the variation in the measured candidate masses. It also implicitly assumes that all of the candidate BNS events are indeed astrophysical. Such an assumption has further implications for their classifications -- for example, under the assumption that S190426c is of astrophysical origin, it is found to more likely be an NSBH merger \citep{GCN24411}. Our na\"{i}ve estimate therefore provides a rough guide as to the volumetric BNS merger rate, but should be treated with caution.

\begin{figure}
\centering\includegraphics[width = \columnwidth]{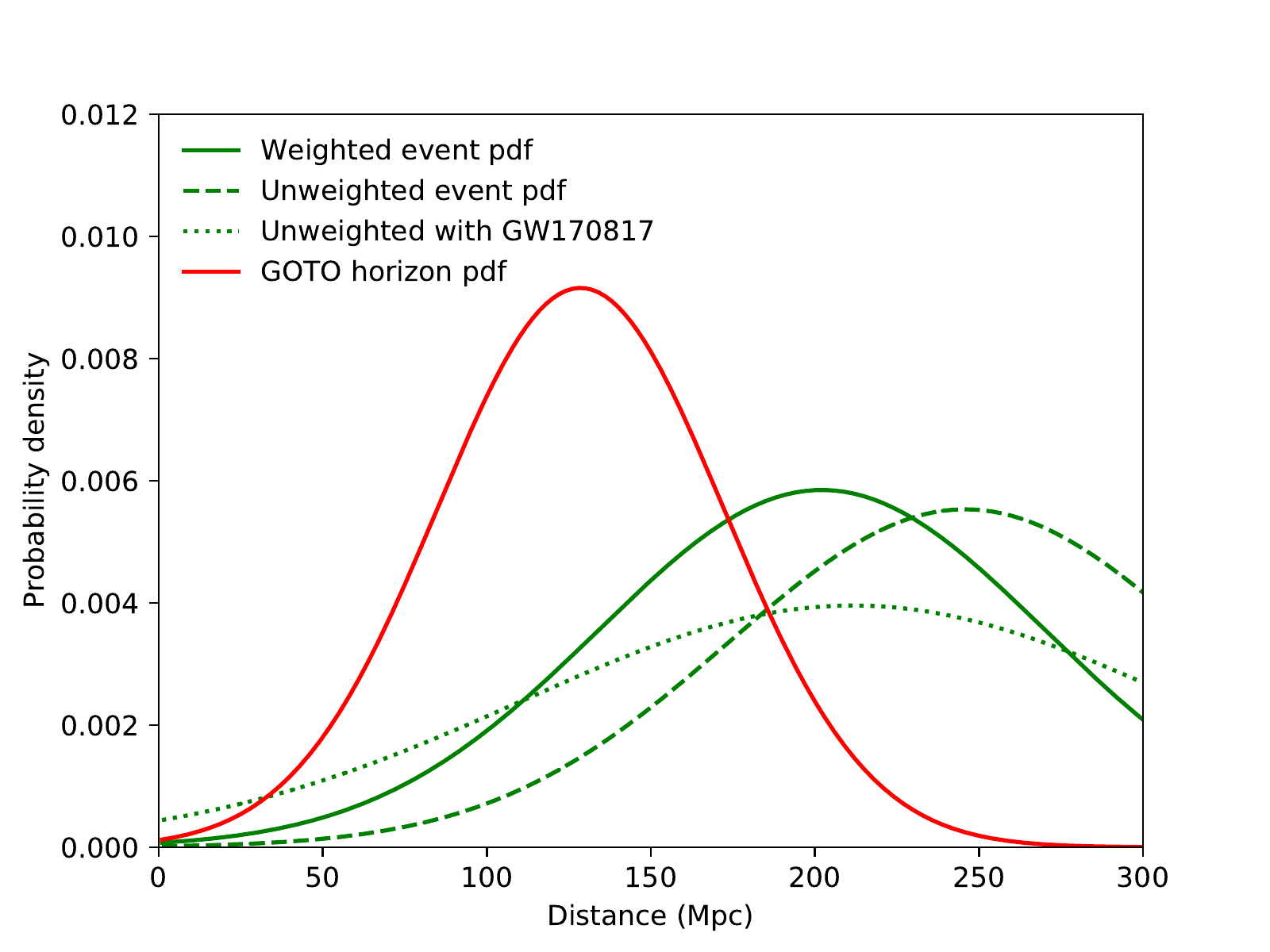}
\caption{The probability density function (PDF) of the horizons achieved for a Bazin model during the 29 GOTO-4 LVC follow-up campaigns (red). These are compared to the weighted PDF (weights $ = 1/\sigma_{\rm dist}^2$) of the 5 BNS merger events during this time (green, solid), their unweighted PDF (dashed), and the unweighted PDF when the distance to GW170817 \citep[41~Mpc;][]{Hjorth17} is included (dotted). The weighted PDF including GW170817 is not shown because the distance errors for this event are far smaller than the other 5 due to its EM detection.}
\label{fig:pdfs}
\end{figure}

Fig.~\ref{fig:pdfs} shows the distance distribution of events classified as BNS by the LVC alongside the distribution of observable horizons (assuming a Bazin model) achieved by GOTO-4 during LVC follow-up. We make 100,000 draws from the GOTO-4 horizon probability density function (PDF), and compare it to 100,000 draws from both the weighted event distance PDF (the weighting for each BNS is $1/\sigma_{\rm dist}^2$, see Table~\ref{tab:sample}), and the unweighted event distance PDF including GW170817. Assuming each draw is a unique event with unique follow-up, we find that GOTO-4 reaches the depth required for a detection in $18.3$ ($22.5$) per cent of cases for the weighted (unweighted + GW170817) distributions.

\subsection{Factors that Influence Performance}\label{sec:conditions}
We further investigate events with unusually high or low observable horizons. The average maximum horizon achieved was $126 \pm 40$~Mpc (mean$\pm 1\sigma$ standard deviation in the sample). The LVC superevents with the worst observable horizons are S190915ak, S190510g and S190421ar. At 15~Mpc, S190915ak is the worst by far, and the explanation is an unfortunate combination of high airmass (2) and a bright moon ($91.5$ per cent illumination). These poor conditions were coupled with a single observing epoch taken almost a full day after peak, when our test model had decayed almost a full magnitude from maximum brightness. For S190510g, we attained a maximum horizon of 57~Mpc. The airmass was $1.78$ though the illumination was low at just 30 per cent. In this case, the largest detrimental factor was actually observing too early; all epochs were within $0.1$ days of the trigger, preceding the rapid rise to peak of the test model (Fig.~\ref{fig:test_sources}). The contemporaneous model was two magnitudes below peak. S190421ar had a maximum observable horizon of 66~Mpc, an airmass of $1.88$ a fairly high moon illumination of 77 per cent, and also endured high winds, which were on average 21~km/h during exposure. This event also suffered from higher extinction along the line of sight compared to the other campaigns, with a median value of $A_V \approx 0.2$ mags; more than double the typical median. Furthermore, this event was only observed two days after trigger; two magnitudes below the model peak.

At the high end of the observable horizon distribution, only S190425z is more than $1\sigma$ above the mean. The maximum observable horizon achieved in this case was 227~Mpc. The moon illumination was 51 per cent, and the airmass was low, at $1.27$. While these conditions are clearly better than the three aforementioned cases, perhaps the biggest contributor to the distant horizon is the fact that the bulk of the observations were taken right at the peak of our model flux. The next highest horizon in the sample was 168~Mpc for S190706ai, which had an airmass of $1.79$ and a moon illumination of 34 per cent. These conditions are very similar to S190510g, with the key difference being that S190706ai was observed much closer to the model peak (though not as close as 190425z).

The depth of $m_{\rm lim}$ is clearly affected by the weather, wind shake, airmass, moon brightness, and the telescope optics, among other things. However, the indication from our analysis of the O3a campaign is that while observing conditions do play a role in determining from how far away we can expect to detect a KN (and can render a campaign entirely unconstraining to our models, as in S190915ak), the timing of the observation with respect to peak flux is in fact the dominant variable. This is reflected in the mean $m_{\rm lim}$ values; while S190915ak clearly experienced very poor observing conditions, with $m_{\rm lim} = 15.0$, the other four events discussed have mean $m_{\rm lim}$ ranging between $19.3$ (S190421ar) and $20.0$ (S190425z). Despite this, their observing horizons vary greatly due to the differing proximity of the observations to the model peak. For our Bazin model based on AT2017gfo, observing as close to $0.5$ days after trigger as possible is therefore highly desirable. One major caveat to this is that KN evolution at early times ($\sim$ a few hours) is largely unknown due to a lack of observations. Our model at these times is therefore not well constrained (see Section~\ref{sec:kne}), and hence the poor recovery performance of events with only very early observations may be pessimistic. Additionally, AT2017gfo is only one (well studied) event. The indication from cosmological GRBs is that KNe show considerable diversity in their emission \citep{Fong17,gompertz2018diversity,Ascenzi19,Rossi20}, meaning that the model employed here may be too optimistic/pessimistic in its peak magnitude, or evolve faster/slower than any given future event. AT2017gfo is fainter than all but one sGRB KN candidate, but in several cases non-detections that probe deeper than the AT2017gfo models imply that there is room for a fainter, undetected population \citep{gompertz2018diversity,Pandey19}.

\subsection{Future Prospects}
In addition to the factors discussed in Section~\ref{sec:conditions}, improvements in the limiting magnitudes can be made with a simple increase in exposure time. There are two obvious ways that the horizons presented in this paper can be improved: stacking the existing exposures, and/or increasing the exposure time for future follow-up campaigns. The former method could potentially yield depth increases of $\sim 0.85$ magnitudes, since our stacking gains scale as $\approx 2.5 \log\sqrt{N}$ where $N$ is the number of stacks and our mean duplication factor is $4.8$. The second method to increase depth will happen naturally as GOTO approaches design specifications; adding more telescopes means that the footprint of an individual tile/pointing becomes larger, and hence the GW probability regions can be tiled faster. The result is more time for repeat visits and/or longer exposure times, as well as more recent reference tiles from a higher survey cadence. Additionally, as the GW detector network expands to design sensitivity, the localisation precision of GW triggers (in particular nearby BNS mergers) is expected to improve \citep{Abbott20}, meaning that it may no longer be necessary to tile many thousands of square degrees. The most recent estimates \citep{Abbott20} for the fourth LIGO/Virgo run (O4), which begins in January 2022 and will include the Kamioka Gravitational-wave Detector (KAGRA) suggests that the BNS (BBH) localisation precision will improve from a median $270^{+34}_{-30}$ ($280^{+30}_{-23}$) square degree 90 per cent credible interval in O3 to a median $33^{+5}_{-5}$ ($41^{+7}_{-6}$) square degree 90 per cent credible interval during O4. These factors will improve the depth of the observations on average. In particular, the flexible design of GOTO enables it to point both mounts at a single tile to improve depth once the localisations have improved to the point where breadth is no longer an issue.

\begin{figure}
\centering\includegraphics[width = \columnwidth]{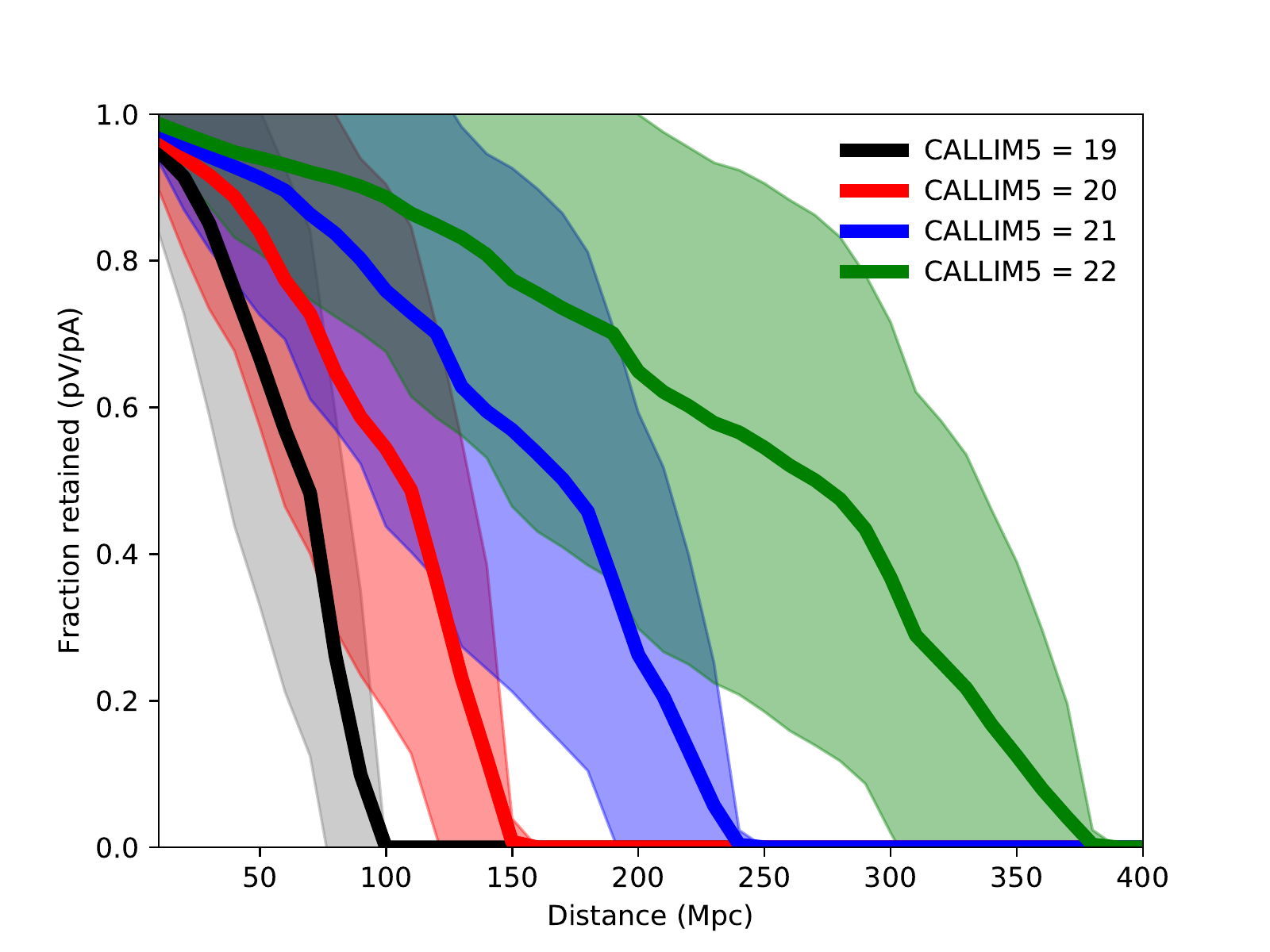}
\caption{The mean fraction of 2D probability coverage that is retained with distance for our sample using an assumed $m_{\rm lim}$ for all tiles. "CALLIM5" is the GOTO fits header keyword for $m_{\rm lim}$. In addition to greater depth overall, we expect more visits to produce a more "box-like" plot, with flatter sections out to greater distances, followed by faster drop offs. This is because more repeats along a given line of sight will provide deeper alternatives to poor observing epochs with shallow horizons, which create the sloping curvature seen here.}
\label{fig:callim5}
\end{figure}

With regards to the sensitivity required, the expected BNS range during O4 \citep{Abbott20} is 160 -- 190~Mpc (aLIGO), 90 -- 120~Mpc (AdV) and 25 -- 130~Mpc (KAGRA). For comparison, Fig.~\ref{fig:callim5} shows GOTO's expected probability recovery fraction with distance in increments of $m_{\rm lim}$. This incorporates the real observing conditions and follow-up schedule for each of the 29 events in O3a; it differs from Fig.~\ref{fig:pVpA} only in that $m_{\rm lim}$ is held constant for every tile. We find that attaining $m_{\rm lim}$ of magnitude 22 (Fig.~\ref{fig:callim5}, green line) will provide the range necessary for the predicted O4 BNS distribution for all but the most distant events. This is attainable for GOTO with an increase in exposure time or further image stacking (see e.g. Steeghs et al., in prep).

The mean response time of $8.79$ hours from the time that the LVC preliminary notice is received is impeded by notices that are received during the La Palma day. The forthcoming upgrade to add a southern node in Australia will further improve GOTO's response time by increasing the window in which one of the facilities can respond (for overlapping latitudes), as well as greatly increasing the range of observable declinations.

\section{Conclusions}\label{sec:conclusions}

We find that even in its 4-telescope prototype configuration, GOTO covered $70.3$ per cent of the \emph{available} probability region on average when following up LVC superevents. The mean area covered was $732$ square degrees, but up to $2667$ square degrees were covered in a single campaign. Over 500 square degrees were observed during 15 of the campaigns, including more than 1000 square degrees in 6 of them. In cases of well-timed events that fell in unconstrained tiles, GOTO-4 began observations less than a minute after receiving the LVC alert. On average, GOTO began observations $8.79$ hours after receiving an alert ($9.90$ hours after the GW trigger). Despite no detections of GW-EM counterpart candidates in O3a, the telescope therefore comfortably fulfilled its role as a wide-field GW follow-up facility.

During the first half of O3, a typical GOTO observing campaign covered $45.3$ per cent of the LVC probability map, and could have unearthed an AT2017gfo-like KN up to 126~Mpc away. GOTO-4 achieves a maximum depth sufficient to recover our KN test source in one-in-five follow-up attempts for a distance drawn from the BNS distance distribution in O3a. However, we find that due to their distance, it is not possible to place model-constraining limits on EM emission from the distant (> 250~Mpc) population of BBH mergers detected by the LVC unless they house on-axis gamma-ray bursts; GOTO-4 was able to cover almost 10 per cent of the total probability \emph{volume} on average for an on-axis GRB model, and rule one out on three occasions. We also note that future GOTO GW follow-up strategy would benefit from returning to a candidate field at $\sim 5$ -- $10$~days post-trigger in order to better constrain the possibility of an off-axis GRB.

Based on our findings, the primary focus for GOTO should now be improving depth. The average duplication factor of $4.8$ visits per LVC probability pixel and the recent upgrade to 8 telescopes will help to achieve this goal. We find that reaching a $5\sigma$ limit of 22 magnitudes will provide KN coverage of almost the entire BNS volume probed by GW interferometers during O4 (though this `volume' will be anisotropic, and many mergers will be detected at or near the horizon, where most of the volume is). This is expected to be comfortably within reach once GOTO attains its full 2 node, 32 telescope configuration. Efforts to improve depth will be aided by improved localisations from the expanding network of GW detectors, which will allow longer GOTO exposures and/or more repeated visits, and the option to point both mounts at a single target. The second GOTO node in Australia will also allow very rapid response in a greater fraction of cases, as well as enable follow-up of triggers whose fields lie predominantly at southern latitudes. The versatile design of GOTO means that it can evolve as the GW detectors do.

\section*{Acknowledgements}

We thank the anonymous referee for helpful comments that improved the manuscript.

BPG, AJL and KW have received funding from the European Research Council (ERC) under the European Union's Horizon 2020 research and innovation programme (grant agreement no 725246, TEDE, PI Levan). 

DS, KU and JL acknowledge support from the STFC via grants ST/T007184/1, ST/T003103/1 and ST/P000495/1.

Armagh Observatory \& Planetarium is core funded by the NI Government through the Dept. for Communities.

EP acknowledges funding from the Spanish Ministry of Economics and Competitiveness through project PGC2018-098153-B-C31.

RPB, MRK and DMS acknowledge support from the ERC under the European Union's Horizon 2020 research and innovation programme (grant agreement No. 715051; Spiders).

RLCS, POB and RAJEF acknowledge funding from STFC.

The Gravitational-wave Optical Transient Observer (GOTO) project acknowledges the support of the Monash-Warwick Alliance; Warwick University; Monash University; Sheffield University; the University of Leicester; Armagh Observatory \& Planetarium; the National Astronomical Research Institute of Thailand (NARIT); the University of Turku; Portsmouth University; and the Instituto de Astrof\'{i}sica de Canarias (IAC).

\section*{Data Availability}

Data products will be available as part of planned GOTO public data releases.

\bibliographystyle{mnras}
\bibliography{BIB}

\begin{thebibliography}{}
\makeatletter
\relax
\def\mn@urlcharsother{\let\do\@makeother \do\$\do\&\do\#\do\^\do\_\do\%\do\~}
\def\mn@doi{\begingroup\mn@urlcharsother \@ifnextchar [ {\mn@doi@}
  {\mn@doi@[]}}
\def\mn@doi@[#1]#2{\def\@tempa{#1}\ifx\@tempa\@empty \href
  {http://dx.doi.org/#2} {doi:#2}\else \href {http://dx.doi.org/#2} {#1}\fi
  \endgroup}
\def\mn@eprint#1#2{\mn@eprint@#1:#2::\@nil}
\def\mn@eprint@arXiv#1{\href {http://arxiv.org/abs/#1} {{\tt arXiv:#1}}}
\def\mn@eprint@dblp#1{\href {http://dblp.uni-trier.de/rec/bibtex/#1.xml}
  {dblp:#1}}
\def\mn@eprint@#1:#2:#3:#4\@nil{\def\@tempa {#1}\def\@tempb {#2}\def\@tempc
  {#3}\ifx \@tempc \@empty \let \@tempc \@tempb \let \@tempb \@tempa \fi \ifx
  \@tempb \@empty \def\@tempb {arXiv}\fi \@ifundefined
  {mn@eprint@\@tempb}{\@tempb:\@tempc}{\expandafter \expandafter \csname
  mn@eprint@\@tempb\endcsname \expandafter{\@tempc}}}

\bibitem[\protect\citeauthoryear{{Aasi} et~al.,}{{Aasi} et~al.}{2013}]{Aasi13}
{Aasi} J.,  et~al., 2013, \mn@doi [\prd] {10.1103/PhysRevD.88.062001}, \href
  {https://ui.adsabs.harvard.edu/abs/2013PhRvD..88f2001A} {88, 062001}

\bibitem[\protect\citeauthoryear{{Abadie} et~al.,}{{Abadie}
  et~al.}{2010}]{Abadie10}
{Abadie} J.,  et~al., 2010, \mn@doi [Classical and Quantum Gravity]
  {10.1088/0264-9381/27/17/173001}, \href
  {https://ui.adsabs.harvard.edu/abs/2010CQGra..27q3001A} {27, 173001}

\bibitem[\protect\citeauthoryear{Abbott et~al.,}{Abbott
  et~al.}{2016}]{abbott2016localization}
Abbott B.~P.,  et~al., 2016, The Astrophysical journal letters, 826, L13

\bibitem[\protect\citeauthoryear{{Abbott} et~al.,}{{Abbott}
  et~al.}{2017a}]{Abbott17b}
{Abbott} B.~P.,  et~al., 2017a, \mn@doi [Physical Review Letters]
  {10.1103/PhysRevLett.119.161101}, \href
  {http://adsabs.harvard.edu/abs/2017PhRvL.119p1101A} {119, 161101}

\bibitem[\protect\citeauthoryear{{Abbott} et~al.,}{{Abbott}
  et~al.}{2017b}]{abbott2017multi}
{Abbott} B.~P.,  et~al., 2017b, \mn@doi [\apjl] {10.3847/2041-8213/aa91c9},
  \href {http://adsabs.harvard.edu/abs/2017ApJ...848L..12A} {848, L12}

\bibitem[\protect\citeauthoryear{Abbott et~al.,}{Abbott
  et~al.}{2019}]{Abbott19}
Abbott B.~P.,  et~al., 2019, \mn@doi [Phys. Rev. X]
  {10.1103/PhysRevX.9.031040}, 9, 031040

\bibitem[\protect\citeauthoryear{{Abbott} et~al.,}{{Abbott}
  et~al.}{2020a}]{Abbott20}
{Abbott} B.~P.,  et~al., 2020a, \mn@doi [Living Reviews in Relativity]
  {10.1007/s41114-018-0012-9}, \href
  {https://ui.adsabs.harvard.edu/abs/2018LRR....21....3A} {21, 3}

\bibitem[\protect\citeauthoryear{{Abbott} et~al.,}{{Abbott}
  et~al.}{2020b}]{Abbott20b}
{Abbott} B.~P.,  et~al., 2020b, \mn@doi [\apjl] {10.3847/2041-8213/ab75f5},
  \href {https://ui.adsabs.harvard.edu/abs/2020ApJ...892L...3A} {892, L3}

\bibitem[\protect\citeauthoryear{{Ackley} et~al.,}{{Ackley}
  et~al.}{2020}]{Ackley20}
{Ackley} K.,  et~al., 2020, arXiv e-prints, \href
  {https://ui.adsabs.harvard.edu/abs/2020arXiv200201950A} {p. arXiv:2002.01950}

\bibitem[\protect\citeauthoryear{{Anand} et~al.,}{{Anand}
  et~al.}{2020}]{Anand20}
{Anand} S.,  et~al., 2020, arXiv e-prints, \href
  {https://ui.adsabs.harvard.edu/abs/2020arXiv200305516A} {p. arXiv:2003.05516}

\bibitem[\protect\citeauthoryear{{Andreoni} et~al.,}{{Andreoni}
  et~al.}{2017}]{Andreoni17}
{Andreoni} I.,  et~al., 2017, \mn@doi [\pasa] {10.1017/pasa.2017.65}, \href
  {https://ui.adsabs.harvard.edu/abs/2017PASA...34...69A} {34, e069}

\bibitem[\protect\citeauthoryear{{Andreoni} et~al.,}{{Andreoni}
  et~al.}{2020}]{Andreoni20}
{Andreoni} I.,  et~al., 2020, \mn@doi [\apj] {10.3847/1538-4357/ab6a1b}, \href
  {https://ui.adsabs.harvard.edu/abs/2020ApJ...890..131A} {890, 131}

\bibitem[\protect\citeauthoryear{{Antier} et~al.,}{{Antier}
  et~al.}{2020a}]{Antier20b}
{Antier} S.,  et~al., 2020a, arXiv e-prints, \href
  {https://ui.adsabs.harvard.edu/abs/2020arXiv200404277A} {p. arXiv:2004.04277}

\bibitem[\protect\citeauthoryear{{Antier} et~al.,}{{Antier}
  et~al.}{2020b}]{Antier20}
{Antier} S.,  et~al., 2020b, \mn@doi [\mnras] {10.1093/mnras/stz3142}, \href
  {https://ui.adsabs.harvard.edu/abs/2020MNRAS.492.3904A} {492, 3904}

\bibitem[\protect\citeauthoryear{{Arcavi} et~al.,}{{Arcavi}
  et~al.}{2017}]{Arcavi17}
{Arcavi} I.,  et~al., 2017, \mn@doi [\nat] {10.1038/nature24291}, \href
  {https://ui.adsabs.harvard.edu/abs/2017Natur.551...64A} {551, 64}

\bibitem[\protect\citeauthoryear{{Ascenzi} et~al.,}{{Ascenzi}
  et~al.}{2019}]{Ascenzi19}
{Ascenzi} S.,  et~al., 2019, \mn@doi [\mnras] {10.1093/mnras/stz891}, \href
  {https://ui.adsabs.harvard.edu/abs/2019MNRAS.486..672A} {486, 672}

\bibitem[\protect\citeauthoryear{{Barbieri}, {Salafia}, {Perego}, {Colpi}  \&
  {Ghirlanda}}{{Barbieri} et~al.}{2019}]{Barbieri19}
{Barbieri} C.,  {Salafia} O.~S.,  {Perego} A.,  {Colpi} M.,   {Ghirlanda} G.,
  2019, \mn@doi [\aap] {10.1051/0004-6361/201935443}, \href
  {https://ui.adsabs.harvard.edu/abs/2019A&A...625A.152B} {625, A152}

\bibitem[\protect\citeauthoryear{{Barnes} \& {Kasen}}{{Barnes} \&
  {Kasen}}{2013}]{Barnes13}
{Barnes} J.,  {Kasen} D.,  2013, \mn@doi [\apj] {10.1088/0004-637X/775/1/18},
  \href {http://adsabs.harvard.edu/abs/2013ApJ...775...18B} {775, 18}

\bibitem[\protect\citeauthoryear{Bazin et~al.,}{Bazin
  et~al.}{2011}]{bazin2011photometric}
Bazin G.,  et~al., 2011, Astronomy \& Astrophysics, 534, A43

\bibitem[\protect\citeauthoryear{Becker}{Becker}{2015}]{becker2015hotpants}
Becker A.,  2015, Astrophysics Source Code Library

\bibitem[\protect\citeauthoryear{Belczynski, Perna, Bulik, Kalogera, Ivanova
  \& Lamb}{Belczynski et~al.}{2006}]{belczynski2006study}
Belczynski K.,  Perna R.,  Bulik T.,  Kalogera V.,  Ivanova N.,   Lamb D.~Q.,
  2006, The Astrophysical Journal, 648, 1110

\bibitem[\protect\citeauthoryear{{Bellm} et~al.,}{{Bellm}
  et~al.}{2019}]{Bellm19}
{Bellm} E.~C.,  et~al., 2019, \mn@doi [\pasp] {10.1088/1538-3873/aaecbe}, \href
  {https://ui.adsabs.harvard.edu/abs/2019PASP..131a8002B} {131, 018002}

\bibitem[\protect\citeauthoryear{{Berger}, {Fong}  \& {Chornock}}{{Berger}
  et~al.}{2013}]{Berger13}
{Berger} E.,  {Fong} W.,   {Chornock} R.,  2013, \mn@doi [\apjl]
  {10.1088/2041-8205/774/2/L23}, \href
  {https://ui.adsabs.harvard.edu/abs/2013ApJ...774L..23B} {774, L23}

\bibitem[\protect\citeauthoryear{{Blanchard} et~al.,}{{Blanchard}
  et~al.}{2017}]{Blanchard17}
{Blanchard} P.~K.,  et~al., 2017, \mn@doi [\apjl] {10.3847/2041-8213/aa9055},
  \href {http://adsabs.harvard.edu/abs/2017ApJ...848L..22B} {848, L22}

\bibitem[\protect\citeauthoryear{{Blandford} \& {McKee}}{{Blandford} \&
  {McKee}}{1976}]{Blandford76}
{Blandford} R.~D.,  {McKee} C.~F.,  1976, \mn@doi [Physics of Fluids]
  {10.1063/1.861619}, \href {http://adsabs.harvard.edu/abs/1976PhFl...19.1130B}
  {19, 1130}

\bibitem[\protect\citeauthoryear{{Blinnikov}, {Novikov}, {Perevodchikova}  \&
  {Polnarev}}{{Blinnikov} et~al.}{1984}]{Blinnikov84}
{Blinnikov} S.~I.,  {Novikov} I.~D.,  {Perevodchikova} T.~V.,   {Polnarev}
  A.~G.,  1984, Soviet Astronomy Letters, \href
  {https://ui.adsabs.harvard.edu/abs/1984SvAL...10..177B} {10, 177}

\bibitem[\protect\citeauthoryear{Bloemen et~al.,}{Bloemen
  et~al.}{2016}]{Bloeman16}
Bloemen S.,  et~al., 2016, in Hall H.~J.,  Gilmozzi R.,   Marshall H.~K.,  eds,
   Vol. 9906, Ground-based and Airborne Telescopes VI. SPIE, pp 2118 -- 2126,
  \mn@doi{10.1117/12.2232522}, \url {https://doi.org/10.1117/12.2232522}

\bibitem[\protect\citeauthoryear{{Bloom} et~al.,}{{Bloom}
  et~al.}{2012}]{Bloom12}
{Bloom} J.~S.,  et~al., 2012, \mn@doi [\pasp] {10.1086/668468}, \href
  {https://ui.adsabs.harvard.edu/abs/2012PASP..124.1175B} {124, 1175}

\bibitem[\protect\citeauthoryear{{Bo{\"e}r}}{{Bo{\"e}r}}{2001}]{Boer01}
{Bo{\"e}r} M.,  2001, \mn@doi [Astronomische Nachrichten]
  {10.1002/1521-3994(200112)322:5/6<343::AID-ASNA343>3.0.CO;2-2}, \href
  {https://ui.adsabs.harvard.edu/abs/2001AN....322..343B} {322, 343}

\bibitem[\protect\citeauthoryear{{Capaccioli} \& {Schipani}}{{Capaccioli} \&
  {Schipani}}{2011}]{Capaccioli11}
{Capaccioli} M.,  {Schipani} P.,  2011, The Messenger, \href
  {https://ui.adsabs.harvard.edu/abs/2011Msngr.146....2C} {146, 2}

\bibitem[\protect\citeauthoryear{{Chang} \& {Murray}}{{Chang} \&
  {Murray}}{2018}]{Chang18}
{Chang} P.,  {Murray} N.,  2018, \mn@doi [\mnras] {10.1093/mnrasl/slx186},
  \href {http://adsabs.harvard.edu/abs/2018MNRAS.474L..12C} {474, L12}

\bibitem[\protect\citeauthoryear{{Chornock} et~al.,}{{Chornock}
  et~al.}{2017}]{Chornock17}
{Chornock} R.,  et~al., 2017, \mn@doi [\apjl] {10.3847/2041-8213/aa905c}, \href
  {http://adsabs.harvard.edu/abs/2017ApJ...848L..19C} {848, L19}

\bibitem[\protect\citeauthoryear{Connaughton et~al.,}{Connaughton
  et~al.}{2016}]{connaughton2016fermi}
Connaughton V.,  et~al., 2016, The Astrophysical Journal Letters, 826, L6

\bibitem[\protect\citeauthoryear{{Coughlin} et~al.,}{{Coughlin}
  et~al.}{2019}]{Coughlin19}
{Coughlin} M.~W.,  et~al., 2019, \mn@doi [\apjl] {10.3847/2041-8213/ab4ad8},
  \href {https://ui.adsabs.harvard.edu/abs/2019ApJ...885L..19C} {885, L19}

\bibitem[\protect\citeauthoryear{{Coughlin} et~al.,}{{Coughlin}
  et~al.}{2020}]{Coughlin20}
{Coughlin} M.~W.,  et~al., 2020, \mn@doi [\mnras] {10.1093/mnras/stz3457},
  \href {https://ui.adsabs.harvard.edu/abs/2020MNRAS.492..863C} {492, 863}

\bibitem[\protect\citeauthoryear{{Coulter} et~al.,}{{Coulter}
  et~al.}{2017}]{Coulter17a}
{Coulter} D.~A.,  et~al., 2017, \mn@doi [Science] {10.1126/science.aap9811},
  \href {http://adsabs.harvard.edu/abs/2017Sci...358.1556C} {358, 1556}

\bibitem[\protect\citeauthoryear{{Covino} et~al.,}{{Covino}
  et~al.}{2017}]{Covino17}
{Covino} S.,  et~al., 2017, \mn@doi [Nature Astronomy]
  {10.1038/s41550-017-0285-z}, \href
  {http://adsabs.harvard.edu/abs/2017NatAs...1..791C} {1, 791}

\bibitem[\protect\citeauthoryear{{Cowperthwaite} et~al.,}{{Cowperthwaite}
  et~al.}{2017}]{Cowperthwaite17}
{Cowperthwaite} P.~S.,  et~al., 2017, \mn@doi [\apjl]
  {10.3847/2041-8213/aa8fc7}, \href
  {http://adsabs.harvard.edu/abs/2017ApJ...848L..17C} {848, L17}

\bibitem[\protect\citeauthoryear{{D'Avanzo} et~al.,}{{D'Avanzo}
  et~al.}{2018}]{DAvanzo18}
{D'Avanzo} P.,  et~al., 2018, \mn@doi [\aap] {10.1051/0004-6361/201832664},
  \href {http://adsabs.harvard.edu/abs/2018A%26A...613L...1D} {613, L1}

\bibitem[\protect\citeauthoryear{D{\'a}lya et~al.,}{D{\'a}lya
  et~al.}{2018}]{dalya2018glade}
D{\'a}lya G.,  et~al., 2018, Monthly Notices of the Royal Astronomical Society,
  479, 2374

\bibitem[\protect\citeauthoryear{{Dark Energy Survey Collaboration}
  et~al.,}{{Dark Energy Survey Collaboration} et~al.}{2016}]{DES16}
{Dark Energy Survey Collaboration} et~al., 2016, \mn@doi [\mnras]
  {10.1093/mnras/stw641}, \href
  {https://ui.adsabs.harvard.edu/abs/2016MNRAS.460.1270D} {460, 1270}

\bibitem[\protect\citeauthoryear{De~Mink \& King}{De~Mink \&
  King}{2017}]{de2017electromagnetic}
De~Mink S.,  King A.,  2017, The Astrophysical Journal Letters, 839, L7

\bibitem[\protect\citeauthoryear{{D{\'\i}az} et~al.,}{{D{\'\i}az}
  et~al.}{2017}]{Diaz17}
{D{\'\i}az} M.~C.,  et~al., 2017, \mn@doi [\apjl] {10.3847/2041-8213/aa9060},
  \href {https://ui.adsabs.harvard.edu/abs/2017ApJ...848L..29D} {848, L29}

\bibitem[\protect\citeauthoryear{{Dobie} et~al.,}{{Dobie}
  et~al.}{2019}]{Dobie19}
{Dobie} D.,  et~al., 2019, \mn@doi [\apjl] {10.3847/2041-8213/ab59db}, \href
  {https://ui.adsabs.harvard.edu/abs/2019ApJ...887L..13D} {887, L13}

\bibitem[\protect\citeauthoryear{{Drout} et~al.,}{{Drout}
  et~al.}{2017}]{Drout17}
{Drout} M.~R.,  et~al., 2017, \mn@doi [Science] {10.1126/science.aaq0049},
  \href {http://adsabs.harvard.edu/abs/2017Sci...358.1570D} {358, 1570}

\bibitem[\protect\citeauthoryear{{Dyer}, {Dhillon}, {Littlefair}, {Steeghs},
  {Ulaczyk}, {Chote}, {Galloway}  \& {Rol}}{{Dyer}
  et~al.}{2018}]{dyer2018telescope}
{Dyer} M.~J.,  {Dhillon} V.~S.,  {Littlefair} S.,  {Steeghs} D.,  {Ulaczyk} K.,
   {Chote} P.,  {Galloway} D.,   {Rol} E.,  2018, in \procspie. p. 107040C
  (\mn@eprint {arXiv} {1807.01614}), \mn@doi{10.1117/12.2311865}

\bibitem[\protect\citeauthoryear{{Eichler}, {Livio}, {Piran}  \&
  {Schramm}}{{Eichler} et~al.}{1989}]{Eichler89}
{Eichler} D.,  {Livio} M.,  {Piran} T.,   {Schramm} D.~N.,  1989, \mn@doi
  [\nat] {10.1038/340126a0}, \href
  {http://adsabs.harvard.edu/abs/1989Natur.340..126E} {340, 126}

\bibitem[\protect\citeauthoryear{{Evans} et~al.,}{{Evans}
  et~al.}{2017}]{Evans17}
{Evans} P.~A.,  et~al., 2017, \mn@doi [Science] {10.1126/science.aap9580},
  \href {http://adsabs.harvard.edu/abs/2017Sci...358.1565E} {358, 1565}

\bibitem[\protect\citeauthoryear{{Fong} \& {Berger}}{{Fong} \&
  {Berger}}{2013}]{Fong13}
{Fong} W.,  {Berger} E.,  2013, \mn@doi [\apj] {10.1088/0004-637X/776/1/18},
  \href {https://ui.adsabs.harvard.edu/abs/2013ApJ...776...18F} {776, 18}

\bibitem[\protect\citeauthoryear{{Fong}, {Berger}, {Margutti}  \&
  {Zauderer}}{{Fong} et~al.}{2015}]{Fong15}
{Fong} W.,  {Berger} E.,  {Margutti} R.,   {Zauderer} B.~A.,  2015, \mn@doi
  [\apj] {10.1088/0004-637X/815/2/102}, \href
  {http://adsabs.harvard.edu/abs/2015ApJ...815..102F} {815, 102}

\bibitem[\protect\citeauthoryear{{Fong} et~al.,}{{Fong} et~al.}{2017}]{Fong17}
{Fong} W.,  et~al., 2017, \mn@doi [\apjl] {10.3847/2041-8213/aa9018}, \href
  {https://ui.adsabs.harvard.edu/abs/2017ApJ...848L..23F} {848, L23}

\bibitem[\protect\citeauthoryear{{Fong} et~al.,}{{Fong} et~al.}{2019}]{Fong19}
{Fong} W.,  et~al., 2019, \mn@doi [\apjl] {10.3847/2041-8213/ab3d9e}, \href
  {https://ui.adsabs.harvard.edu/abs/2019ApJ...883L...1F} {883, L1}

\bibitem[\protect\citeauthoryear{{Foucart}, {Hinderer}  \&
  {Nissanke}}{{Foucart} et~al.}{2018}]{Foucart18}
{Foucart} F.,  {Hinderer} T.,   {Nissanke} S.,  2018, \mn@doi [\prd]
  {10.1103/PhysRevD.98.081501}, \href
  {https://ui.adsabs.harvard.edu/abs/2018PhRvD..98h1501F} {98, 081501}

\bibitem[\protect\citeauthoryear{{Foucart}, {Duez}, {Kidder}, {Nissanke},
  {Pfeiffer}  \& {Scheel}}{{Foucart} et~al.}{2019}]{Foucart19}
{Foucart} F.,  {Duez} M.~D.,  {Kidder} L.~E.,  {Nissanke} S.~M.,  {Pfeiffer}
  H.~P.,   {Scheel} M.~A.,  2019, \mn@doi [\prd] {10.1103/PhysRevD.99.103025},
  \href {https://ui.adsabs.harvard.edu/abs/2019PhRvD..99j3025F} {99, 103025}

\bibitem[\protect\citeauthoryear{{Freiburghaus}, {Rosswog}  \&
  {Thielemann}}{{Freiburghaus} et~al.}{1999}]{Freiburghaus99}
{Freiburghaus} C.,  {Rosswog} S.,   {Thielemann} F.-K.,  1999, \mn@doi [\apjl]
  {10.1086/312343}, \href {http://adsabs.harvard.edu/abs/1999ApJ...525L.121F}
  {525, L121}

\bibitem[\protect\citeauthoryear{{Goldstein} et~al.,}{{Goldstein}
  et~al.}{2017}]{Goldstein17}
{Goldstein} A.,  et~al., 2017, \mn@doi [\apjl] {10.3847/2041-8213/aa8f41},
  \href {http://adsabs.harvard.edu/abs/2017ApJ...848L..14G} {848, L14}

\bibitem[\protect\citeauthoryear{{Goldstein} et~al.,}{{Goldstein}
  et~al.}{2019}]{Goldstein19}
{Goldstein} D.~A.,  et~al., 2019, \mn@doi [\apjl] {10.3847/2041-8213/ab3046},
  \href {https://ui.adsabs.harvard.edu/abs/2019ApJ...881L...7G} {881, L7}

\bibitem[\protect\citeauthoryear{{Gomez} et~al.,}{{Gomez}
  et~al.}{2019}]{Gomez19}
{Gomez} S.,  et~al., 2019, \mn@doi [\apjl] {10.3847/2041-8213/ab4ad5}, \href
  {https://ui.adsabs.harvard.edu/abs/2019ApJ...884L..55G} {884, L55}

\bibitem[\protect\citeauthoryear{Gompertz, van~der Horst, O'Brien, Wynn  \&
  Wiersema}{Gompertz et~al.}{2015}]{gompertz2015broad}
Gompertz B.~P.,  van~der Horst A.,  O'Brien P.~T.,  Wynn G.~A.,   Wiersema K.,
  2015, Monthly Notices of the Royal Astronomical Society, 448, 629

\bibitem[\protect\citeauthoryear{Gompertz et~al.,}{Gompertz
  et~al.}{2018}]{gompertz2018diversity}
Gompertz B.,  et~al., 2018, The Astrophysical Journal, 860, 62

\bibitem[\protect\citeauthoryear{{Gompertz}, {Levan}  \& {Tanvir}}{{Gompertz}
  et~al.}{2020}]{Gompertz20}
{Gompertz} B.~P.,  {Levan} A.~J.,   {Tanvir} N.~R.,  2020, arXiv e-prints,
  \href {https://ui.adsabs.harvard.edu/abs/2020arXiv200108706G} {p.
  arXiv:2001.08706}

\bibitem[\protect\citeauthoryear{{G{\'o}rski}, {Hivon}, {Banday}, {Wand elt},
  {Hansen}, {Reinecke}  \& {Bartelmann}}{{G{\'o}rski} et~al.}{2005}]{Gorski05}
{G{\'o}rski} K.~M.,  {Hivon} E.,  {Banday} A.~J.,  {Wand elt} B.~D.,  {Hansen}
  F.~K.,  {Reinecke} M.,   {Bartelmann} M.,  2005, \mn@doi [\apj]
  {10.1086/427976}, \href
  {https://ui.adsabs.harvard.edu/abs/2005ApJ...622..759G} {622, 759}

\bibitem[\protect\citeauthoryear{{Granot}, {Gill}, {Guetta}  \& {De
  Colle}}{{Granot} et~al.}{2018}]{Granot18}
{Granot} J.,  {Gill} R.,  {Guetta} D.,   {De Colle} F.,  2018, \mn@doi [\mnras]
  {10.1093/mnras/sty2308}, \href
  {https://ui.adsabs.harvard.edu/abs/2018MNRAS.481.1597G} {481, 1597}

\bibitem[\protect\citeauthoryear{{Guillochon}, {Parrent}, {Kelley}  \&
  {Margutti}}{{Guillochon} et~al.}{2017}]{Guillochon17}
{Guillochon} J.,  {Parrent} J.,  {Kelley} L.~Z.,   {Margutti} R.,  2017,
  \mn@doi [\apj] {10.3847/1538-4357/835/1/64}, \href
  {https://ui.adsabs.harvard.edu/abs/2017ApJ...835...64G} {835, 64}

\bibitem[\protect\citeauthoryear{{Haggard}, {Nynka}, {Ruan}, {Kalogera},
  {Cenko}, {Evans}  \& {Kennea}}{{Haggard} et~al.}{2017}]{Haggard17}
{Haggard} D.,  {Nynka} M.,  {Ruan} J.~J.,  {Kalogera} V.,  {Cenko} S.~B.,
  {Evans} P.,   {Kennea} J.~A.,  2017, \mn@doi [\apjl]
  {10.3847/2041-8213/aa8ede}, \href
  {http://adsabs.harvard.edu/abs/2017ApJ...848L..25H} {848, L25}

\bibitem[\protect\citeauthoryear{{Hallinan} et~al.,}{{Hallinan}
  et~al.}{2017}]{Hallinan17}
{Hallinan} G.,  et~al., 2017, \mn@doi [Science] {10.1126/science.aap9855},
  \href {http://adsabs.harvard.edu/abs/2017Sci...358.1579H} {358, 1579}

\bibitem[\protect\citeauthoryear{Henden, Templeton, Terrell, Smith, Levine  \&
  Welch}{Henden et~al.}{2016}]{henden2016vizier}
Henden A.,  Templeton M.,  Terrell D.,  Smith T.,  Levine S.,   Welch D.,
  2016, VizieR Online Data Catalog, 2336

\bibitem[\protect\citeauthoryear{{Hjorth} et~al.,}{{Hjorth}
  et~al.}{2017}]{Hjorth17}
{Hjorth} J.,  et~al., 2017, \mn@doi [\apjl] {10.3847/2041-8213/aa9110}, \href
  {https://ui.adsabs.harvard.edu/abs/2017ApJ...848L..31H} {848, L31}

\bibitem[\protect\citeauthoryear{{Hosseinzadeh} et~al.,}{{Hosseinzadeh}
  et~al.}{2019}]{Hosseinzadeh19}
{Hosseinzadeh} G.,  et~al., 2019, \mn@doi [\apjl] {10.3847/2041-8213/ab271c},
  \href {https://ui.adsabs.harvard.edu/abs/2019ApJ...880L...4H} {880, L4}

\bibitem[\protect\citeauthoryear{Janiuk, Bejger, Charzy{\'n}ski  \&
  Sukova}{Janiuk et~al.}{2017}]{janiuk2017possible}
Janiuk A.,  Bejger M.,  Charzy{\'n}ski S.,   Sukova P.,  2017, New Astronomy,
  51, 7

\bibitem[\protect\citeauthoryear{{Jin} et~al.,}{{Jin} et~al.}{2016}]{Jin16}
{Jin} Z.-P.,  et~al., 2016, \mn@doi [Nature Communications]
  {10.1038/ncomms12898}, \href
  {http://adsabs.harvard.edu/abs/2016NatCo...712898J} {7, 12898}

\bibitem[\protect\citeauthoryear{{Jin} et~al.,}{{Jin} et~al.}{2018}]{Jin18}
{Jin} Z.-P.,  et~al., 2018, \mn@doi [\apj] {10.3847/1538-4357/aab76d}, \href
  {http://adsabs.harvard.edu/abs/2018ApJ...857..128J} {857, 128}

\bibitem[\protect\citeauthoryear{{Jin}, {Covino}, {Liao}, {Li}, {D'Avanzo},
  {Fan}  \& {Wei}}{{Jin} et~al.}{2020}]{Jin20}
{Jin} Z.-P.,  {Covino} S.,  {Liao} N.-H.,  {Li} X.,  {D'Avanzo} P.,  {Fan}
  Y.-Z.,   {Wei} D.-M.,  2020, \mn@doi [Nature Astronomy]
  {10.1038/s41550-019-0892-y}, \href
  {https://ui.adsabs.harvard.edu/abs/2020NatAs...4...77J} {4, 77}

\bibitem[\protect\citeauthoryear{{Kaiser} et~al.,}{{Kaiser}
  et~al.}{2010}]{Kaiser10}
{Kaiser} N.,  et~al., 2010, in \procspie. p. 77330E, \mn@doi{10.1117/12.859188}

\bibitem[\protect\citeauthoryear{{Kasliwal}, {Korobkin}, {Lau}, {Wollaeger}  \&
  {Fryer}}{{Kasliwal} et~al.}{2017}]{Kasliwal17}
{Kasliwal} M.~M.,  {Korobkin} O.,  {Lau} R.~M.,  {Wollaeger} R.,   {Fryer}
  C.~L.,  2017, \mn@doi [\apjl] {10.3847/2041-8213/aa799d}, \href
  {http://adsabs.harvard.edu/abs/2017ApJ...843L..34K} {843, L34}

\bibitem[\protect\citeauthoryear{{Kawaguchi}, {Kyutoku}, {Nakano}, {Okawa},
  {Shibata}  \& {Taniguchi}}{{Kawaguchi} et~al.}{2015}]{Kawaguchi15}
{Kawaguchi} K.,  {Kyutoku} K.,  {Nakano} H.,  {Okawa} H.,  {Shibata} M.,
  {Taniguchi} K.,  2015, \mn@doi [\prd] {10.1103/PhysRevD.92.024014}, \href
  {https://ui.adsabs.harvard.edu/abs/2015PhRvD..92b4014K} {92, 024014}

\bibitem[\protect\citeauthoryear{{Kawaguchi}, {Kyutoku}, {Shibata}  \&
  {Tanaka}}{{Kawaguchi} et~al.}{2016}]{Kawaguchi16}
{Kawaguchi} K.,  {Kyutoku} K.,  {Shibata} M.,   {Tanaka} M.,  2016, \mn@doi
  [\apj] {10.3847/0004-637X/825/1/52}, \href
  {http://adsabs.harvard.edu/abs/2016ApJ...825...52K} {825, 52}

\bibitem[\protect\citeauthoryear{{Kawaguchi}, {Shibata}  \&
  {Tanaka}}{{Kawaguchi} et~al.}{2020}]{Kawaguchi20}
{Kawaguchi} K.,  {Shibata} M.,   {Tanaka} M.,  2020, \mn@doi [\apj]
  {10.3847/1538-4357/ab61f6}, \href
  {https://ui.adsabs.harvard.edu/abs/2020ApJ...889..171K} {889, 171}

\bibitem[\protect\citeauthoryear{{Kim} et~al.,}{{Kim} et~al.}{2016}]{Kim16}
{Kim} S.-L.,  et~al., 2016, \mn@doi [Journal of Korean Astronomical Society]
  {10.5303/JKAS.2016.49.1.037}, \href
  {https://ui.adsabs.harvard.edu/abs/2016JKAS...49...37K} {49, 37}

\bibitem[\protect\citeauthoryear{{Kim} et~al.,}{{Kim} et~al.}{2017}]{Kim17}
{Kim} S.,  et~al., 2017, \mn@doi [\apjl] {10.3847/2041-8213/aa970b}, \href
  {http://adsabs.harvard.edu/abs/2017ApJ...850L..21K} {850, L21}

\bibitem[\protect\citeauthoryear{Kool et~al.,}{Kool et~al.}{2019}]{kool19}
Kool E.,  et~al., 2019, GCN, 25616, 1

\bibitem[\protect\citeauthoryear{{Kouveliotou}, {Meegan}, {Fishman}, {Bhat},
  {Briggs}, {Koshut}, {Paciesas}  \& {Pendleton}}{{Kouveliotou}
  et~al.}{1993}]{Kouveliotou93}
{Kouveliotou} C.,  {Meegan} C.~A.,  {Fishman} G.~J.,  {Bhat} N.~P.,  {Briggs}
  M.~S.,  {Koshut} T.~M.,  {Paciesas} W.~S.,   {Pendleton} G.~N.,  1993,
  \mn@doi [\apjl] {10.1086/186969}, \href
  {http://adsabs.harvard.edu/abs/1993ApJ...413L.101K} {413, L101}

\bibitem[\protect\citeauthoryear{{Kyutoku}, {Ioka}, {Okawa}, {Shibata}  \&
  {Taniguchi}}{{Kyutoku} et~al.}{2015}]{Kyutoku15}
{Kyutoku} K.,  {Ioka} K.,  {Okawa} H.,  {Shibata} M.,   {Taniguchi} K.,  2015,
  \mn@doi [\prd] {10.1103/PhysRevD.92.044028}, \href
  {https://ui.adsabs.harvard.edu/abs/2015PhRvD..92d4028K} {92, 044028}

\bibitem[\protect\citeauthoryear{{LIGO Scientific Collaboration} \& {Virgo
  Collaboration}}{{LIGO Scientific Collaboration} \& {Virgo
  Collaboration}}{2019}]{LSC19}
{LIGO Scientific Collaboration} {Virgo Collaboration} 2019, GRB Coordinates
  Network, \href {https://ui.adsabs.harvard.edu/abs/2019GCN.25333....1L}
  {25333, 1}

\bibitem[\protect\citeauthoryear{{Lamb} \& {Kobayashi}}{{Lamb} \&
  {Kobayashi}}{2018}]{Lamb18}
{Lamb} G.~P.,  {Kobayashi} S.,  2018, \mn@doi [\mnras] {10.1093/mnras/sty1108},
  \href {http://adsabs.harvard.edu/abs/2018MNRAS.478..733L} {478, 733}

\bibitem[\protect\citeauthoryear{{Lamb} et~al.,}{{Lamb} et~al.}{2019a}]{Lamb19}
{Lamb} G.~P.,  et~al., 2019a, \mn@doi [\apjl] {10.3847/2041-8213/aaf96b}, \href
  {http://adsabs.harvard.edu/abs/2019ApJ...870L..15L} {870, L15}

\bibitem[\protect\citeauthoryear{{Lamb} et~al.,}{{Lamb}
  et~al.}{2019b}]{Lamb19b}
{Lamb} G.~P.,  et~al., 2019b, \mn@doi [\apj] {10.3847/1538-4357/ab38bb}, \href
  {https://ui.adsabs.harvard.edu/abs/2019ApJ...883...48L} {883, 48}

\bibitem[\protect\citeauthoryear{Lang, Hogg, Mierle, Blanton  \& Roweis}{Lang
  et~al.}{2010}]{lang2010astrometry}
Lang D.,  Hogg D.~W.,  Mierle K.,  Blanton M.,   Roweis S.,  2010, The
  astronomical journal, 139, 1782

\bibitem[\protect\citeauthoryear{{Lattimer} \& {Schramm}}{{Lattimer} \&
  {Schramm}}{1974}]{Lattimer74}
{Lattimer} J.~M.,  {Schramm} D.~N.,  1974, \mn@doi [\apjl] {10.1086/181612},
  \href {http://adsabs.harvard.edu/abs/1974ApJ...192L.145L} {192, L145}

\bibitem[\protect\citeauthoryear{{Lazzati}, {Perna}, {Morsony},
  {L{\'o}pez-C{\'a}mara}, {Cantiello}, {Ciolfi}, {giacomazzo}  \&
  {Workman}}{{Lazzati} et~al.}{2017}]{Lazzati17}
{Lazzati} D.,  {Perna} R.,  {Morsony} B.~J.,  {L{\'o}pez-C{\'a}mara} D.,
  {Cantiello} M.,  {Ciolfi} R.,  {giacomazzo} B.,   {Workman} J.~C.,  2017,
  preprint, \href {http://adsabs.harvard.edu/abs/2017arXiv171203237L} {}
  (\mn@eprint {arXiv} {1712.03237})

\bibitem[\protect\citeauthoryear{{Levan} et~al.,}{{Levan}
  et~al.}{2017}]{Levan17}
{Levan} A.~J.,  et~al., 2017, \mn@doi [\apjl] {10.3847/2041-8213/aa905f}, \href
  {http://adsabs.harvard.edu/abs/2017ApJ...848L..28L} {848, L28}

\bibitem[\protect\citeauthoryear{{Li} \& {Paczy{\'n}ski}}{{Li} \&
  {Paczy{\'n}ski}}{1998}]{Li98}
{Li} L.-X.,  {Paczy{\'n}ski} B.,  1998, \mn@doi [\apjl] {10.1086/311680}, \href
  {http://adsabs.harvard.edu/abs/1998ApJ...507L..59L} {507, L59}

\bibitem[\protect\citeauthoryear{{Ligo Collaboration} \& {VIRGO
  Collaboration}}{{Ligo Collaboration} \& {VIRGO
  Collaboration}}{2019a}]{LVCGCNS190901ap}
{Ligo Collaboration} {VIRGO Collaboration} 2019a, GRB Coordinates Network,
  \href {https://gcn.gsfc.nasa.gov/gcn3/25606.gcn3} {25606, 1}

\bibitem[\protect\citeauthoryear{{Ligo Collaboration} \& {VIRGO
  Collaboration}}{{Ligo Collaboration} \& {VIRGO
  Collaboration}}{2019b}]{LVCGCNS190915ak}
{Ligo Collaboration} {VIRGO Collaboration} 2019b, GRB Coordinates Network,
  \href {https://gcn.gsfc.nasa.gov/gcn3/25753.gcn3} {25753, 1}

\bibitem[\protect\citeauthoryear{{Ligo Collaboration} \& {VIRGO
  Collaboration}}{{Ligo Collaboration} \& {VIRGO
  Collaboration}}{2019c}]{LVCGCNS190924h}
{Ligo Collaboration} {VIRGO Collaboration} 2019c, GRB Coordinates Network,
  \href {https://gcn.gsfc.nasa.gov/gcn3/25829.gcn3} {25829, 1}

\bibitem[\protect\citeauthoryear{{Ligo Collaboration} \& {VIRGO
  Collaboration}}{{Ligo Collaboration} \& {VIRGO
  Collaboration}}{2019d}]{LVCGCNS190930s}
{Ligo Collaboration} {VIRGO Collaboration} 2019d, GRB Coordinates Network,
  \href {https://gcn.gsfc.nasa.gov/gcn3/25871.gcn3} {25871, 1}

\bibitem[\protect\citeauthoryear{{Ligo Scientific Collaboration} \& {VIRGO
  Collaboration}}{{Ligo Scientific Collaboration} \& {VIRGO
  Collaboration}}{2019a}]{LVCGCNS190408an}
{Ligo Scientific Collaboration} {VIRGO Collaboration} 2019a, GRB Coordinates
  Network, \href {https://ui.adsabs.harvard.edu/abs/2019GCN.24069....1L}
  {24069, 1}

\bibitem[\protect\citeauthoryear{{Ligo Scientific Collaboration} \& {VIRGO
  Collaboration}}{{Ligo Scientific Collaboration} \& {VIRGO
  Collaboration}}{2019b}]{LVCGCNS190412m}
{Ligo Scientific Collaboration} {VIRGO Collaboration} 2019b, GRB Coordinates
  Network, \href {https://ui.adsabs.harvard.edu/abs/2019GCN.24098....1L}
  {24098, 1}

\bibitem[\protect\citeauthoryear{{Ligo Scientific Collaboration} \& {VIRGO
  Collaboration}}{{Ligo Scientific Collaboration} \& {VIRGO
  Collaboration}}{2019c}]{LVC2GCNS190421ar}
{Ligo Scientific Collaboration} {VIRGO Collaboration} 2019c, GRB Coordinates
  Network, \href {https://ui.adsabs.harvard.edu/abs/2019GCN.24141....1L}
  {24141, 1}

\bibitem[\protect\citeauthoryear{{Ligo Scientific Collaboration} \& {VIRGO
  Collaboration}}{{Ligo Scientific Collaboration} \& {VIRGO
  Collaboration}}{2019d}]{LVCGCNS190425z}
{Ligo Scientific Collaboration} {VIRGO Collaboration} 2019d, GRB Coordinates
  Network, \href {https://ui.adsabs.harvard.edu/abs/2019GCN.24168....1L}
  {24168, 1}

\bibitem[\protect\citeauthoryear{{Ligo Scientific Collaboration} \& {VIRGO
  Collaboration}}{{Ligo Scientific Collaboration} \& {VIRGO
  Collaboration}}{2019e}]{LVCGCNS190426c}
{Ligo Scientific Collaboration} {VIRGO Collaboration} 2019e, GRB Coordinates
  Network, \href {https://ui.adsabs.harvard.edu/abs/2019GCN.24237....1L}
  {24237, 1}

\bibitem[\protect\citeauthoryear{{Ligo Scientific Collaboration} \& {VIRGO
  Collaboration}}{{Ligo Scientific Collaboration} \& {VIRGO
  Collaboration}}{2019f}]{GCN24411}
{Ligo Scientific Collaboration} {VIRGO Collaboration} 2019f, GRB Coordinates
  Network, \href {https://ui.adsabs.harvard.edu/abs/2019GCN.24411....1L}
  {24411, 1}

\bibitem[\protect\citeauthoryear{{Ligo Scientific Collaboration} \& {VIRGO
  Collaboration}}{{Ligo Scientific Collaboration} \& {VIRGO
  Collaboration}}{2019g}]{LVCGCNS190510g}
{Ligo Scientific Collaboration} {VIRGO Collaboration} 2019g, GRB Coordinates
  Network, \href {https://gcn.gsfc.nasa.gov/gcn3/24442.gcn3} {24442, 1}

\bibitem[\protect\citeauthoryear{{Ligo Scientific Collaboration} \& {VIRGO
  Collaboration}}{{Ligo Scientific Collaboration} \& {VIRGO
  Collaboration}}{2019h}]{LVCGCNS190512at}
{Ligo Scientific Collaboration} {VIRGO Collaboration} 2019h, GRB Coordinates
  Network, \href {https://gcn.gsfc.nasa.gov/gcn3/24503.gcn3} {24503, 1}

\bibitem[\protect\citeauthoryear{{Ligo Scientific Collaboration} \& {VIRGO
  Collaboration}}{{Ligo Scientific Collaboration} \& {VIRGO
  Collaboration}}{2019i}]{LVCGCNS190513bm}
{Ligo Scientific Collaboration} {VIRGO Collaboration} 2019i, GRB Coordinates
  Network, \href {https://gcn.gsfc.nasa.gov/gcn3/24522.gcn3} {24522, 1}

\bibitem[\protect\citeauthoryear{{Ligo Scientific Collaboration} \& {VIRGO
  Collaboration}}{{Ligo Scientific Collaboration} \& {VIRGO
  Collaboration}}{2019j}]{LVCGCNS190517h}
{Ligo Scientific Collaboration} {VIRGO Collaboration} 2019j, GRB Coordinates
  Network, \href {https://gcn.gsfc.nasa.gov/gcn3/24570.gcn3} {24570, 1}

\bibitem[\protect\citeauthoryear{{Ligo Scientific Collaboration} \& {VIRGO
  Collaboration}}{{Ligo Scientific Collaboration} \& {VIRGO
  Collaboration}}{2019k}]{LVCGCNS190519bj}
{Ligo Scientific Collaboration} {VIRGO Collaboration} 2019k, GRB Coordinates
  Network, \href {https://gcn.gsfc.nasa.gov/gcn3/24598.gcn3} {24598, 1}

\bibitem[\protect\citeauthoryear{{Ligo Scientific Collaboration} \& {VIRGO
  Collaboration}}{{Ligo Scientific Collaboration} \& {VIRGO
  Collaboration}}{2019l}]{LVCGCNS190521g}
{Ligo Scientific Collaboration} {VIRGO Collaboration} 2019l, GRB Coordinates
  Network, \href {https://gcn.gsfc.nasa.gov/gcn3/24621.gcn3} {24621, 1}

\bibitem[\protect\citeauthoryear{{Ligo Scientific Collaboration} \& {VIRGO
  Collaboration}}{{Ligo Scientific Collaboration} \& {VIRGO
  Collaboration}}{2019m}]{LVCGCNS190521r}
{Ligo Scientific Collaboration} {VIRGO Collaboration} 2019m, GRB Coordinates
  Network, \href {https://gcn.gsfc.nasa.gov/gcn3/24632.gcn3} {24632, 1}

\bibitem[\protect\citeauthoryear{{Ligo Scientific Collaboration} \& {VIRGO
  Collaboration}}{{Ligo Scientific Collaboration} \& {VIRGO
  Collaboration}}{2019n}]{LVCGCNS190630ag}
{Ligo Scientific Collaboration} {VIRGO Collaboration} 2019n, GRB Coordinates
  Network, \href {https://gcn.gsfc.nasa.gov/gcn3/24922.gcn3} {24922, 1}

\bibitem[\protect\citeauthoryear{{Ligo Scientific Collaboration} \& {VIRGO
  Collaboration}}{{Ligo Scientific Collaboration} \& {VIRGO
  Collaboration}}{2019o}]{LVCGCNS190706ai}
{Ligo Scientific Collaboration} {VIRGO Collaboration} 2019o, GRB Coordinates
  Network, \href {https://gcn.gsfc.nasa.gov/gcn3/24998.gcn3} {24998, 1}

\bibitem[\protect\citeauthoryear{{Ligo Scientific Collaboration} \& {VIRGO
  Collaboration}}{{Ligo Scientific Collaboration} \& {VIRGO
  Collaboration}}{2019p}]{LVCGCNS190707q}
{Ligo Scientific Collaboration} {VIRGO Collaboration} 2019p, GRB Coordinates
  Network, \href {https://gcn.gsfc.nasa.gov/gcn3/25012.gcn3} {25012, 1}

\bibitem[\protect\citeauthoryear{{Ligo Scientific Collaboration} \& {VIRGO
  Collaboration}}{{Ligo Scientific Collaboration} \& {VIRGO
  Collaboration}}{2019q}]{LVCGCNS190718y}
{Ligo Scientific Collaboration} {VIRGO Collaboration} 2019q, GRB Coordinates
  Network, \href {https://gcn.gsfc.nasa.gov/gcn3/25087.gcn3} {25087, 1}

\bibitem[\protect\citeauthoryear{{Ligo Scientific Collaboration} \& {VIRGO
  Collaboration}}{{Ligo Scientific Collaboration} \& {VIRGO
  Collaboration}}{2019r}]{LVCGCNS190720a}
{Ligo Scientific Collaboration} {VIRGO Collaboration} 2019r, GRB Coordinates
  Network, \href {https://gcn.gsfc.nasa.gov/gcn3/25115.gcn3} {25115, 1}

\bibitem[\protect\citeauthoryear{{Ligo Scientific Collaboration} \& {VIRGO
  Collaboration}}{{Ligo Scientific Collaboration} \& {VIRGO
  Collaboration}}{2019s}]{LVCGCNS190727h}
{Ligo Scientific Collaboration} {VIRGO Collaboration} 2019s, GRB Coordinates
  Network, \href {https://gcn.gsfc.nasa.gov/gcn3/25164.gcn3} {25164, 1}

\bibitem[\protect\citeauthoryear{{Ligo Scientific Collaboration} \& {VIRGO
  Collaboration}}{{Ligo Scientific Collaboration} \& {VIRGO
  Collaboration}}{2019t}]{LVCGCNS190728q}
{Ligo Scientific Collaboration} {VIRGO Collaboration} 2019t, GRB Coordinates
  Network, \href {https://gcn.gsfc.nasa.gov/gcn3/25187.gcn3} {25187, 1}

\bibitem[\protect\citeauthoryear{{Ligo Scientific Collaboration} \& {VIRGO
  Collaboration}}{{Ligo Scientific Collaboration} \& {VIRGO
  Collaboration}}{2019u}]{LVCGCNS190814bv}
{Ligo Scientific Collaboration} {VIRGO Collaboration} 2019u, GRB Coordinates
  Network, \href {https://gcn.gsfc.nasa.gov/gcn3/25324.gcn3} {25324, 1}

\bibitem[\protect\citeauthoryear{{Ligo Scientific Collaboration} \& {VIRGO
  Collaboration}}{{Ligo Scientific Collaboration} \& {VIRGO
  Collaboration}}{2019v}]{LVCGCNS190828l}
{Ligo Scientific Collaboration} {VIRGO Collaboration} 2019v, GRB Coordinates
  Network, \href {https://gcn.gsfc.nasa.gov/gcn3/25497.gcn3} {25497, 1}

\bibitem[\protect\citeauthoryear{{Ligo Scientific Collaboration} \& {VIRGO
  Collaboration}}{{Ligo Scientific Collaboration} \& {VIRGO
  Collaboration}}{2019w}]{LVCGCNS190828j}
{Ligo Scientific Collaboration} {VIRGO Collaboration} 2019w, GRB Coordinates
  Network, \href {https://gcn.gsfc.nasa.gov/gcn3/25503.gcn3} {25503, 1}

\bibitem[\protect\citeauthoryear{{Ligo Scientific Collaboration} \& {VIRGO
  Collaboration}}{{Ligo Scientific Collaboration} \& {VIRGO
  Collaboration}}{2019x}]{LVCGCNS190910d}
{Ligo Scientific Collaboration} {VIRGO Collaboration} 2019x, GRB Coordinates
  Network, \href {https://gcn.gsfc.nasa.gov/gcn3/25695.gcn3} {25695, 1}

\bibitem[\protect\citeauthoryear{{Ligo Scientific Collaboration} \& {VIRGO
  Collaboration}}{{Ligo Scientific Collaboration} \& {VIRGO
  Collaboration}}{2019y}]{LVCGCNS190923y}
{Ligo Scientific Collaboration} {VIRGO Collaboration} 2019y, GRB Coordinates
  Network, \href {https://gcn.gsfc.nasa.gov/gcn3/25814.gcn3} {25814, 1}

\bibitem[\protect\citeauthoryear{{Ligo Scientific Collaboration} \& {VIRGO
  Collaboration}}{{Ligo Scientific Collaboration} \& {VIRGO
  Collaboration}}{2019z}]{LVCGCNS190930t}
{Ligo Scientific Collaboration} {VIRGO Collaboration} 2019z, GRB Coordinates
  Network, \href {https://gcn.gsfc.nasa.gov/gcn3/25876.gcn3} {25876, 1}

\bibitem[\protect\citeauthoryear{{Lipunov} et~al.,}{{Lipunov}
  et~al.}{2010}]{Lipunov10}
{Lipunov} V.,  et~al., 2010, \mn@doi [Advances in Astronomy]
  {10.1155/2010/349171}, \href
  {https://ui.adsabs.harvard.edu/abs/2010AdAst2010E..30L} {2010, 349171}

\bibitem[\protect\citeauthoryear{{Lipunov} et~al.,}{{Lipunov}
  et~al.}{2017}]{Lipunov17}
{Lipunov} V.~M.,  et~al., 2017, \mn@doi [\apjl] {10.3847/2041-8213/aa92c0},
  \href {https://ui.adsabs.harvard.edu/abs/2017ApJ...850L...1L} {850, L1}

\bibitem[\protect\citeauthoryear{Loeb}{Loeb}{2016}]{loeb2016electromagnetic}
Loeb A.,  2016, The Astrophysical Journal Letters, 819, L21

\bibitem[\protect\citeauthoryear{{Lundquist} et~al.,}{{Lundquist}
  et~al.}{2019}]{Lundquist19}
{Lundquist} M.~J.,  et~al., 2019, \mn@doi [\apjl] {10.3847/2041-8213/ab32f2},
  \href {https://ui.adsabs.harvard.edu/abs/2019ApJ...881L..26L} {881, L26}

\bibitem[\protect\citeauthoryear{{Lyman} et~al.,}{{Lyman}
  et~al.}{2018}]{Lyman18}
{Lyman} J.~D.,  et~al., 2018, \mn@doi [Nature Astronomy]
  {10.1038/s41550-018-0511-3}, \href
  {http://adsabs.harvard.edu/abs/2018NatAs.tmp...88L} {}

\bibitem[\protect\citeauthoryear{{Mandel}}{{Mandel}}{2018}]{Mandel18}
{Mandel} I.,  2018, \mn@doi [\apjl] {10.3847/2041-8213/aaa6c1}, \href
  {http://adsabs.harvard.edu/abs/2018ApJ...853L..12M} {853, L12}

\bibitem[\protect\citeauthoryear{{Margutti} et~al.,}{{Margutti}
  et~al.}{2017}]{Margutti17}
{Margutti} R.,  et~al., 2017, \mn@doi [\apjl] {10.3847/2041-8213/aa9057}, \href
  {http://adsabs.harvard.edu/abs/2017ApJ...848L..20M} {848, L20}

\bibitem[\protect\citeauthoryear{{Margutti} et~al.,}{{Margutti}
  et~al.}{2018}]{Margutti18}
{Margutti} R.,  et~al., 2018, \mn@doi [\apjl] {10.3847/2041-8213/aab2ad}, \href
  {http://adsabs.harvard.edu/abs/2018ApJ...856L..18M} {856, L18}

\bibitem[\protect\citeauthoryear{{McKernan} et~al.,}{{McKernan}
  et~al.}{2019}]{McKernan19}
{McKernan} B.,  et~al., 2019, \mn@doi [\apjl] {10.3847/2041-8213/ab4886}, \href
  {https://ui.adsabs.harvard.edu/abs/2019ApJ...884L..50M} {884, L50}

\bibitem[\protect\citeauthoryear{{Metzger}}{{Metzger}}{2017}]{Metzger17}
{Metzger} B.~D.,  2017, \mn@doi [Living Reviews in Relativity]
  {10.1007/s41114-017-0006-z}, \href
  {http://adsabs.harvard.edu/abs/2017LRR....20....3M} {20, 3}

\bibitem[\protect\citeauthoryear{Metzger et~al.,}{Metzger
  et~al.}{2010}]{metzger2010electromagnetic}
Metzger B.,  et~al., 2010, Monthly Notices of the Royal Astronomical Society,
  406, 2650

\bibitem[\protect\citeauthoryear{Moesta, Alic, Rezzolla, Zanotti  \&
  Palenzuela}{Moesta et~al.}{2012}]{moesta2012detectability}
Moesta P.,  Alic D.,  Rezzolla L.,  Zanotti O.,   Palenzuela C.,  2012, The
  Astrophysical Journal Letters, 749, L32

\bibitem[\protect\citeauthoryear{{Mooley} et~al.,}{{Mooley}
  et~al.}{2018}]{Mooley18}
{Mooley} K.~P.,  et~al., 2018, \mn@doi [\apjl] {10.3847/2041-8213/aaeda7},
  \href {https://ui.adsabs.harvard.edu/abs/2018ApJ...868L..11M} {868, L11}

\bibitem[\protect\citeauthoryear{Murase, Kashiyama, M{\'e}sz{\'a}ros, Shoemaker
   \& Senno}{Murase et~al.}{2016}]{murase2016ultrafast}
Murase K.,  Kashiyama K.,  M{\'e}sz{\'a}ros P.,  Shoemaker I.,   Senno N.,
  2016, The Astrophysical Journal Letters, 822, L9

\bibitem[\protect\citeauthoryear{{Narayan}, {Paczynski}  \& {Piran}}{{Narayan}
  et~al.}{1992}]{Narayan92}
{Narayan} R.,  {Paczynski} B.,   {Piran} T.,  1992, \mn@doi [\apjl]
  {10.1086/186493}, \href
  {https://ui.adsabs.harvard.edu/abs/1992ApJ...395L..83N} {395, L83}

\bibitem[\protect\citeauthoryear{{Nicholl} et~al.,}{{Nicholl}
  et~al.}{2017}]{Nicholl17}
{Nicholl} M.,  et~al., 2017, \mn@doi [\apjl] {10.3847/2041-8213/aa9029}, \href
  {http://adsabs.harvard.edu/abs/2017ApJ...848L..18N} {848, L18}

\bibitem[\protect\citeauthoryear{Paczynski}{Paczynski}{1986}]{paczynski1986gamma}
Paczynski B.,  1986, The Astrophysical Journal, 308, L43

\bibitem[\protect\citeauthoryear{Palenzuela, Lehner  \& Yoshida}{Palenzuela
  et~al.}{2010}]{palenzuela2010understanding}
Palenzuela C.,  Lehner L.,   Yoshida S.,  2010, Physical Review D, 81, 084007

\bibitem[\protect\citeauthoryear{{Pandey} et~al.,}{{Pandey}
  et~al.}{2019}]{Pandey19}
{Pandey} S.~B.,  et~al., 2019, \mn@doi [\mnras] {10.1093/mnras/stz530}, \href
  {https://ui.adsabs.harvard.edu/abs/2019MNRAS.485.5294P} {485, 5294}

\bibitem[\protect\citeauthoryear{Perley, Ho, Copperwheat, Collaboration
  et~al.}{Perley et~al.}{2019}]{perley19}
Perley D.,  Ho A.,  Copperwheat C.,  Collaboration G.,   et~al., 2019, GCN,
  25643, 1

\bibitem[\protect\citeauthoryear{Perna, Lazzati  \& Giacomazzo}{Perna
  et~al.}{2016}]{perna2016short}
Perna R.,  Lazzati D.,   Giacomazzo B.,  2016, The Astrophysical Journal
  Letters, 821, L18

\bibitem[\protect\citeauthoryear{{Pian} et~al.,}{{Pian} et~al.}{2017}]{Pian17}
{Pian} E.,  et~al., 2017, \mn@doi [\nat] {10.1038/nature24298}, \href
  {http://adsabs.harvard.edu/abs/2017Natur.551...67P} {551, 67}

\bibitem[\protect\citeauthoryear{{Planck Collaboration} et~al.,}{{Planck
  Collaboration} et~al.}{2018}]{PlanckCollaboration18}
{Planck Collaboration} et~al., 2018, arXiv e-prints, \href
  {https://ui.adsabs.harvard.edu/abs/2018arXiv180706209P} {p. arXiv:1807.06209}

\bibitem[\protect\citeauthoryear{{Pozanenko} et~al.,}{{Pozanenko}
  et~al.}{2018}]{Pozanenko18}
{Pozanenko} A.~S.,  et~al., 2018, \mn@doi [\apjl] {10.3847/2041-8213/aaa2f6},
  \href {https://ui.adsabs.harvard.edu/abs/2018ApJ...852L..30P} {852, L30}

\bibitem[\protect\citeauthoryear{{Rossi} et~al.,}{{Rossi}
  et~al.}{2020}]{Rossi20}
{Rossi} A.,  et~al., 2020, \mn@doi [\mnras] {10.1093/mnras/staa479}, \href
  {https://ui.adsabs.harvard.edu/abs/2020MNRAS.493.3379R} {493, 3379}

\bibitem[\protect\citeauthoryear{{Rosswog}}{{Rosswog}}{2005}]{Rosswog05}
{Rosswog} S.,  2005, \mn@doi [\apj] {10.1086/497062}, \href
  {http://adsabs.harvard.edu/abs/2005ApJ...634.1202R} {634, 1202}

\bibitem[\protect\citeauthoryear{{Rosswog}, {Ramirez-Ruiz}  \&
  {Davies}}{{Rosswog} et~al.}{2003}]{Rosswog03}
{Rosswog} S.,  {Ramirez-Ruiz} E.,   {Davies} M.~B.,  2003, \mn@doi [\mnras]
  {10.1046/j.1365-2966.2003.07032.x}, \href
  {https://ui.adsabs.harvard.edu/abs/2003MNRAS.345.1077R} {345, 1077}

\bibitem[\protect\citeauthoryear{{Sagu{\'e}s Carracedo}, {Bulla}, {Feindt}  \&
  {Goobar}}{{Sagu{\'e}s Carracedo} et~al.}{2020}]{SaguesCarracedo20}
{Sagu{\'e}s Carracedo} A.,  {Bulla} M.,  {Feindt} U.,   {Goobar} A.,  2020,
  arXiv e-prints, \href {https://ui.adsabs.harvard.edu/abs/2020arXiv200406137S}
  {p. arXiv:2004.06137}

\bibitem[\protect\citeauthoryear{{Sari}, {Piran}  \& {Narayan}}{{Sari}
  et~al.}{1998}]{Sari1998spectra}
{Sari} R.,  {Piran} T.,   {Narayan} R.,  1998, \mn@doi [\apjl]
  {10.1086/311269}, \href
  {https://ui.adsabs.harvard.edu/abs/1998ApJ...497L..17S} {497, L17}

\bibitem[\protect\citeauthoryear{{Savchenko} et~al.,}{{Savchenko}
  et~al.}{2017}]{Savchenko17}
{Savchenko} V.,  et~al., 2017, \mn@doi [\apjl] {10.3847/2041-8213/aa8f94},
  \href {http://adsabs.harvard.edu/abs/2017ApJ...848L..15S} {848, L15}

\bibitem[\protect\citeauthoryear{{Schlafly} \& {Finkbeiner}}{{Schlafly} \&
  {Finkbeiner}}{2011}]{Schlafly11}
{Schlafly} E.~F.,  {Finkbeiner} D.~P.,  2011, \mn@doi [\apj]
  {10.1088/0004-637X/737/2/103}, \href
  {https://ui.adsabs.harvard.edu/abs/2011ApJ...737..103S} {737, 103}

\bibitem[\protect\citeauthoryear{{Schlafly}, {Finkbeiner}, {Schlegel},
  {Juri{\'c}}, {Ivezi{\'c}}, {Gibson}, {Knapp}  \& {Weaver}}{{Schlafly}
  et~al.}{2010}]{Schlafly10}
{Schlafly} E.~F.,  {Finkbeiner} D.~P.,  {Schlegel} D.~J.,  {Juri{\'c}} M.,
  {Ivezi{\'c}} {\v{Z}}.,  {Gibson} R.~R.,  {Knapp} G.~R.,   {Weaver} B.~A.,
  2010, \mn@doi [\apj] {10.1088/0004-637X/725/1/1175}, \href
  {https://ui.adsabs.harvard.edu/abs/2010ApJ...725.1175S} {725, 1175}

\bibitem[\protect\citeauthoryear{{Schlegel}, {Finkbeiner}  \&
  {Davis}}{{Schlegel} et~al.}{1998}]{Schlegel98}
{Schlegel} D.~J.,  {Finkbeiner} D.~P.,   {Davis} M.,  1998, \mn@doi [\apj]
  {10.1086/305772}, \href
  {https://ui.adsabs.harvard.edu/abs/1998ApJ...500..525S} {500, 525}

\bibitem[\protect\citeauthoryear{{Shappee} et~al.,}{{Shappee}
  et~al.}{2014}]{Shappee14}
{Shappee} B.,  et~al., 2014, in American Astronomical Society Meeting Abstracts
  \#223. p. 236.03

\bibitem[\protect\citeauthoryear{{Shappee} et~al.,}{{Shappee}
  et~al.}{2017}]{Shappee17}
{Shappee} B.~J.,  et~al., 2017, \mn@doi [Science] {10.1126/science.aaq0186},
  \href {https://ui.adsabs.harvard.edu/abs/2017Sci...358.1574S} {358, 1574}

\bibitem[\protect\citeauthoryear{{Shibata} \& {Taniguchi}}{{Shibata} \&
  {Taniguchi}}{2011}]{Shibata11}
{Shibata} M.,  {Taniguchi} K.,  2011, \mn@doi [Living Reviews in Relativity]
  {10.12942/lrr-2011-6}, \href
  {https://ui.adsabs.harvard.edu/abs/2011LRR....14....6S} {14, 6}

\bibitem[\protect\citeauthoryear{{Singer}}{{Singer}}{2015}]{Singer15}
{Singer} L.~P.,  2015, PhD thesis, California Institute of Technology

\bibitem[\protect\citeauthoryear{{Singer} et~al.,}{{Singer}
  et~al.}{2014}]{Singer14}
{Singer} L.~P.,  et~al., 2014, \mn@doi [\apj] {10.1088/0004-637X/795/2/105},
  \href {https://ui.adsabs.harvard.edu/abs/2014ApJ...795..105S} {795, 105}

\bibitem[\protect\citeauthoryear{{Singer} et~al.,}{{Singer}
  et~al.}{2016}]{Singer16}
{Singer} L.~P.,  et~al., 2016, \mn@doi [\apjs] {10.3847/0067-0049/226/1/10},
  \href {https://ui.adsabs.harvard.edu/abs/2016ApJS..226...10S} {226, 10}

\bibitem[\protect\citeauthoryear{{Smartt} et~al.,}{{Smartt}
  et~al.}{2017}]{Smartt17}
{Smartt} S.~J.,  et~al., 2017, \mn@doi [\nat] {10.1038/nature24303}, \href
  {http://adsabs.harvard.edu/abs/2017Natur.551...75S} {551, 75}

\bibitem[\protect\citeauthoryear{{Soares-Santos} et~al.,}{{Soares-Santos}
  et~al.}{2017}]{Soares-Santos17}
{Soares-Santos} M.,  et~al., 2017, \mn@doi [\apjl] {10.3847/2041-8213/aa9059},
  \href {http://adsabs.harvard.edu/abs/2017ApJ...848L..16S} {848, L16}

\bibitem[\protect\citeauthoryear{{Sutherland} et~al.,}{{Sutherland}
  et~al.}{2015}]{Sutherland15}
{Sutherland} W.,  et~al., 2015, \mn@doi [\aap] {10.1051/0004-6361/201424973},
  \href {https://ui.adsabs.harvard.edu/abs/2015A&A...575A..25S} {575, A25}

\bibitem[\protect\citeauthoryear{{Tanaka} et~al.,}{{Tanaka}
  et~al.}{2018}]{Tanaka18}
{Tanaka} M.,  et~al., 2018, \mn@doi [\apj] {10.3847/1538-4357/aaa0cb}, \href
  {https://ui.adsabs.harvard.edu/abs/2018ApJ...852..109T} {852, 109}

\bibitem[\protect\citeauthoryear{{Tanvir}, {Levan}, {Fruchter}, {Hjorth},
  {Hounsell}, {Wiersema}  \& {Tunnicliffe}}{{Tanvir} et~al.}{2013}]{Tanvir13}
{Tanvir} N.~R.,  {Levan} A.~J.,  {Fruchter} A.~S.,  {Hjorth} J.,  {Hounsell}
  R.~A.,  {Wiersema} K.,   {Tunnicliffe} R.~L.,  2013, \mn@doi [\nat]
  {10.1038/nature12505}, \href
  {https://ui.adsabs.harvard.edu/abs/2013Natur.500..547T} {500, 547}

\bibitem[\protect\citeauthoryear{{Tanvir} et~al.,}{{Tanvir}
  et~al.}{2017}]{Tanvir17}
{Tanvir} N.~R.,  et~al., 2017, \mn@doi [\apjl] {10.3847/2041-8213/aa90b6},
  \href {http://adsabs.harvard.edu/abs/2017ApJ...848L..27T} {848, L27}

\bibitem[\protect\citeauthoryear{{The LIGO Scientific Collaboration}
  et~al.,}{{The LIGO Scientific Collaboration} et~al.}{2020}]{LIGO20}
{The LIGO Scientific Collaboration} et~al., 2020, arXiv e-prints, \href
  {https://ui.adsabs.harvard.edu/abs/2020arXiv200101761T} {p. arXiv:2001.01761}

\bibitem[\protect\citeauthoryear{{Tonry} et~al.,}{{Tonry}
  et~al.}{2018}]{Tonry18}
{Tonry} J.~L.,  et~al., 2018, \mn@doi [\pasp] {10.1088/1538-3873/aabadf}, \href
  {https://ui.adsabs.harvard.edu/abs/2018PASP..130f4505T} {130, 064505}

\bibitem[\protect\citeauthoryear{{Troja}, {King}, {O'Brien}, {Lyons}  \&
  {Cusumano}}{{Troja} et~al.}{2008}]{Troja08}
{Troja} E.,  {King} A.~R.,  {O'Brien} P.~T.,  {Lyons} N.,   {Cusumano} G.,
  2008, \mn@doi [\mnras] {10.1111/j.1745-3933.2007.00421.x}, \href
  {https://ui.adsabs.harvard.edu/abs/2008MNRAS.385L..10T} {385, L10}

\bibitem[\protect\citeauthoryear{{Troja} et~al.,}{{Troja}
  et~al.}{2017}]{Troja17b}
{Troja} E.,  et~al., 2017, \mn@doi [\nat] {10.1038/nature24290}, \href
  {http://adsabs.harvard.edu/abs/2017Natur.551...71T} {551, 71}

\bibitem[\protect\citeauthoryear{{Troja} et~al.,}{{Troja}
  et~al.}{2018a}]{Troja18}
{Troja} E.,  et~al., 2018a, \mn@doi [Nature Communications]
  {10.1038/s41467-018-06558-7}, \href
  {http://adsabs.harvard.edu/abs/2018NatCo...9.4089T} {9, 4089}

\bibitem[\protect\citeauthoryear{{Troja} et~al.,}{{Troja}
  et~al.}{2018b}]{Troja18b}
{Troja} E.,  et~al., 2018b, \mn@doi [\mnras] {10.1093/mnrasl/sly061}, \href
  {http://adsabs.harvard.edu/abs/2018MNRAS.478L..18T} {478, L18}

\bibitem[\protect\citeauthoryear{{Troja} et~al.,}{{Troja}
  et~al.}{2019}]{Troja19}
{Troja} E.,  et~al., 2019, arXiv e-prints, \href
  {https://ui.adsabs.harvard.edu/abs/2019arXiv190501290T} {p. arXiv:1905.01290}

\bibitem[\protect\citeauthoryear{{Valenti} et~al.,}{{Valenti}
  et~al.}{2017}]{Valenti17}
{Valenti} S.,  et~al., 2017, \mn@doi [\apjl] {10.3847/2041-8213/aa8edf}, \href
  {https://ui.adsabs.harvard.edu/abs/2017ApJ...848L..24V} {848, L24}

\bibitem[\protect\citeauthoryear{Veitch et~al.,}{Veitch
  et~al.}{2015}]{veitch15}
Veitch J.,  et~al., 2015, Physical Review D, 91, 042003

\bibitem[\protect\citeauthoryear{{Vieira} et~al.,}{{Vieira}
  et~al.}{2020}]{Vieira20}
{Vieira} N.,  et~al., 2020, arXiv e-prints, \href
  {https://ui.adsabs.harvard.edu/abs/2020arXiv200309437V} {p. arXiv:2003.09437}

\bibitem[\protect\citeauthoryear{{Villar} et~al.,}{{Villar}
  et~al.}{2017}]{Villar17}
{Villar} V.~A.,  et~al., 2017, \mn@doi [\apjl] {10.3847/2041-8213/aa9c84},
  \href {http://adsabs.harvard.edu/abs/2017ApJ...851L..21V} {851, L21}

\bibitem[\protect\citeauthoryear{{Watson} et~al.,}{{Watson}
  et~al.}{2016}]{Watson16}
{Watson} A.~M.,  et~al., 2016, in \procspie. p. 99100G (\mn@eprint {arXiv}
  {1606.00695}), \mn@doi{10.1117/12.2232898}

\bibitem[\protect\citeauthoryear{{Watson} et~al.,}{{Watson}
  et~al.}{2020}]{Watson20}
{Watson} A.~M.,  et~al., 2020, \mn@doi [\mnras] {10.1093/mnras/staa161}, \href
  {https://ui.adsabs.harvard.edu/abs/2020MNRAS.492.5916W} {492, 5916}

\bibitem[\protect\citeauthoryear{{Wells}, {Greisen}  \& {Harten}}{{Wells}
  et~al.}{1981}]{Wells81}
{Wells} D.~C.,  {Greisen} E.~W.,   {Harten} R.~H.,  1981, \aaps, \href
  {https://ui.adsabs.harvard.edu/abs/1981A&AS...44..363W} {44, 363}

\bibitem[\protect\citeauthoryear{{Wu} \& {MacFadyen}}{{Wu} \&
  {MacFadyen}}{2019}]{Wu19}
{Wu} Y.,  {MacFadyen} A.,  2019, \mn@doi [\apjl] {10.3847/2041-8213/ab2fd4},
  \href {https://ui.adsabs.harvard.edu/abs/2019ApJ...880L..23W} {880, L23}

\bibitem[\protect\citeauthoryear{{Yang} et~al.,}{{Yang} et~al.}{2015}]{Yang15}
{Yang} B.,  et~al., 2015, \mn@doi [Nature Communications] {10.1038/ncomms8323},
  \href {http://adsabs.harvard.edu/abs/2015NatCo...6E7323Y} {6, 7323}

\bibitem[\protect\citeauthoryear{{Zhu}, {Yang}, {Liu}, {Huang}, {Zhang}, {Li},
  {Yu}  \& {Gao}}{{Zhu} et~al.}{2020}]{Zhu20}
{Zhu} J.-P.,  {Yang} Y.-P.,  {Liu} L.-D.,  {Huang} Y.,  {Zhang} B.,  {Li} Z.,
  {Yu} Y.-W.,   {Gao} H.,  2020, arXiv e-prints, \href
  {https://ui.adsabs.harvard.edu/abs/2020arXiv200306733Z} {p. arXiv:2003.06733}

\makeatother
\end{thebibliography}

\end{document}